\documentclass[11pt]{article}
\usepackage{epsfig} 
\newcommand{\sect}[1]{ \section{#1} \setcounter{equation}{0} }

\newcommand{\MSbar}{\overline{\mbox{MS}}} 
\newcommand{\MSbars}{\overline{\mbox{\footnotesize{MS}}}} 
\newcommand{\MOMgggg}{\mbox{MOMgggg}}
\newcommand{\MOMggggs}{\mbox{\footnotesize{MOMgggg}}}
\newcommand{\MOMggg}{\mbox{MOMggg}}

\newcommand{\MOMh}{\mbox{MOMh}}

\newcommand{\MOMq}{\mbox{MOMq}}

\newcommand{\Nc}{N_{\!c}}
\newcommand{\Nf}{N_{\!f}}

\begin{document}
\title{Symmetric point quartic gluon vertex and momentum subtraction}
\author{J.A. Gracey, \\ Theoretical Physics Division, \\ 
Department of Mathematical Sciences, \\ University of Liverpool, \\ P.O. Box 
147, \\ Liverpool, \\ L69 3BX, \\ United Kingdom.} 
\date{}
\maketitle 

\vspace{5cm} 
\noindent 
{\bf Abstract.} We compute the full one loop correction to the quartic vertex 
of QCD at the fully symmetric point. This allows us to define a new momentum
subtraction (MOM) scheme in the class of schemes introduced by Celmaster and
Gonsalves. Hence using properties of the renormalization group equation, the 
two loop renormalization group functions for this scheme are given.

\vspace{-15.5cm}
\hspace{13.4cm}
{\bf LTH 1012}

\newpage 

\sect{Introduction.}

In recent years there has been interest in studying the vertex structure of
Quantum Chromodynamics (QCD) which is the quantum field theory corresponding to
the strong nuclear force. The gluon and quark fields which are the constituent
fields and which behave as free entities at high energy are not, however, 
observed in nature. This is due to confinement or infrared slavery. To 
understand why such approximately free particles are restricted to dwell inside
the nucleons of the atomic nucleus requires studying their properties at low 
energy. Theoretically one uses the underlying quantum field theory, QCD, and 
endeavours to ascertain the non-perturbative behaviour of the basic quanta 
interactions. The main computational tools for such investigations are lattice 
gauge theory and Schwinger-Dyson methods. Both techniques complement each 
other. For instance, there has been a large activity in extracting the zero 
momentum limit of the gluon and Faddeev-Popov ghost propagators in the Landau 
gauge. See, for example, various recent articles, \cite{1,2,3,4,5,6,7,8,9,10}. 
While there is a good degree of agreement there is now interest in extending 
such analyses to the $3$-point and $4$-point vertex functions, \cite{11,12}. 
This is a level of difficulty beyond the $2$-point function studies. While 
certain lattice data is available for several vertex functions, 
\cite{13,14,15}, in say the Landau gauge, with general qualitative agreement 
between lattice and Schwinger-Dyson methods, the exploration of these vertex 
functions is not as mature a field as that for propagator analyses. Though it 
should be emphasised that lattice gauge theory studies should improve with 
refined algorithms and supercomputers. While the infrared structure is related 
to confinement issues detailed knowledge is important when considering models 
of hadrons. These physical states of the theory can be examined by modelling 
the dynamics and interactions of the constituent fields via non-perturbative 
knowledge of the vertex functions. 

From the point of view of perturbation theory one can compute the vertex
functions order by order in the loop expansion. Such a field theoretic
programme has been ongoing over several decades. For instance, the one loop
structure of the $3$-point vertices has been studied in \cite{16} at the
fully symmetric point for the external momenta squared and for more general
momentum configurations in \cite{17}. Various extensions have been provided for
a variety of cases. These include two loop corrections with one or more 
external legs on-shell, \cite{18,19,20,21,22}. Moreover, massive quarks have
been included in some cases, \cite{23,24}. More recently the one loop analysis 
of \cite{16} was extended to two loops in \cite{25} with the momentum
subtraction (MOM) scheme renormalization group functions determined for a
linear covariant gauge. To assist model building and to allow Schwinger-Dyson
practitioners to examine whether certain of their truncating approximations or
vertex ans\"{a}tze are reliable, the full off-shell massless vertex functions
for each of the three QCD vertices were determined in \cite{26}. Having 
completed the $3$-point analysis the next stage is to examine the quartic
vertex to complete the set. This is the purpose of this article where we
consider the massless $4$-point gluon vertex at the fully symmetric point, 
which is a non-exceptional point, for a general linear covariant gauge. While 
the extra leg on the Green's function sets it apart from the other vertices its
Feynman rule does not contain a derivative and hence it should not be as 
problematic in that sense as the triple gluon vertex for, say, a lattice 
analysis. We note that, by contrast, in nonlinear covariant gauges one can have
a quartic ghost vertex. Thus, in such gauges one would have to consider 
additional $4$-point vertex computations. Previous work on the quartic gluon 
vertex has not been as intense as the $3$-point ones. Though we note the 
earlier relevant work of \cite{18,27,28,29,30}. For instance, the quartic 
vertex with several on-shell gluons was studied in \cite{27}, while a 
Schwinger-Dyson analysis of the quartic vertex was considered in \cite{28}. 
However, that analysis was not at the fully symmetric point but was for several 
non-exceptional momenta configurations. Moreover, an effective running coupling
constant based on the vertex function was constructed, \cite{28}. An 
alternative recent approach to vertex functions has been provided in 
\cite{29,30} using string theory methods. Based on the triple gluon vertex 
experience of \cite{29}, off-shell quartic vertices have been studied in, for
example, ${\cal N}$~$=$~$4$ supersymmetric Yang-Mills theories in \cite{30}. By 
contrast, in \cite{18} the coupling constant renormalization constant was 
extracted for what was termed the Weinberg scheme. Though unlike what we will 
present here no information on the {\em full} quartic vertex structure at the 
symmetric point was recorded as it was not required for the problem which was 
at hand. Here we will give the full structure. That this is possible is due to 
progress in recent years with computer algebra leading to automatic symbolic 
manipulation programmes as well as with multiloop algorithms. One of the key 
ones has been the provision of the Laporta algorithm, \cite{31}, which has been
implemented in various computer packages including {\sc Reduze}, \cite{32}. 
While we report only on a one loop analysis which demonstrates the viability of
studying the full quartic vertex and thus its potential extension to two loops,
we will also provide a new renormalization scheme in the MOM class of schemes. 
This will be termed the MOMgggg scheme which will have the same ethos of 
definition as the three MOM schemes of \cite{16} which was based on each of the
$3$-point QCD vertices. Although we do not envisage such a scheme overtaking 
the modified minimal subtraction, $\MSbar$, scheme which is the standard 
renormalization scheme, it in some sense completes the set begun in \cite{16}. 
As a byproduct of the scheme definition we will determine its $\Lambda$ 
parameter and evaluate it in relation to $\Lambda_{\MSbars}$. Moreover, once 
the one loop MOMgggg scheme renormalization of the QCD Lagrangian has been 
completed we will use the renormalization group equation to determine all the 
{\em two} loop renormalization group functions without having to complete an 
explicit two loop calculation. Such information will be the bedrock for any 
future computations at this order. Equally our values for the full quartic 
vertex will serve as an independent check if the fully off-shell vertex 
function is found at one loop.

The article is organized as follows. The background for the quartic vertex
computation is discussed in section $2$ together with the relevant group theory
and master integrals. Section $3$ is devoted to explicit results for the
$4$-point function at the fully symmetric point while the MOMgggg scheme is
defined in section $4$. There the two loop MOMgggg renormalization group
functions are recorded. We provide conclusions in section $5$ while an appendix
records the full tensor basis into which we decompose our results.  

\sect{Background.}

In this section we discuss the technical details to our computations. First,
the Green's function we will consider is 
\begin{equation}
\left. \left\langle A^a_\mu(p) A^b_\nu(q) A^c_\sigma(r) A^d_\rho(-p-q-r) 
\right\rangle \right|_{\mbox{\footnotesize{symm}}} ~=~ 
\left. \Sigma^{abcd}_{\mu \nu \sigma \rho}(p,q,r)
\right|_{\mbox{\footnotesize{symm}}} 
\label{quardef}
\end{equation}
where $p$, $q$ and $r$ are the external momentum and we have substituted the 
momentum for the leg with colour label $d$ in terms of the others from 
energy-momentum conservation. The label symm indicates the fully symmetric
point which is defined by 
\begin{equation}
p^2 ~=~ q^2 ~=~ r^2 ~=~ -~ \mu^2 ~~,~~
pq ~=~ pr ~=~ qr ~=~ \frac{1}{3} \mu^2 
\end{equation}
where $\mu$ is a mass scale. Following the same approach as \cite{33,25} we use
the same scale to ensure the coupling constant is dimensionless in
$d$-dimensions as we will regularize dimensionally throughout with
$d$~$=$~$4$~$-$~$2\epsilon$. This external momentum configuration is the same 
as \cite{18} and we note that the Mandelstam variables are then  
\begin{equation}
s ~=~ \frac{1}{2} ( p + q )^2 ~~,~~ t ~=~ \frac{1}{2} ( q + r )^2 ~~,~~
u ~=~ \frac{1}{2} ( p + r )^2
\end{equation}
implying
\begin{equation}
s ~=~ t ~=~ u ~=~ -~ \frac{4}{3} \mu^2 ~.
\end{equation}
In (\ref{quardef}) the right hand side represents the Lorentz and colour
structure. For the $3$-point vertices the colour dependence, at least to low
loop order, involves only one colour tensor which can be readily factored off
so that the focus is purely on the Lorentz component. For the quartic vertex
there is more than one colour structure as will be evident from later
discussions. Therefore, for the moment we indicate how we extract the different
Lorentz channels in (\ref{quardef}). We follow the same method of \cite{25} and
use a projection method. Making no a priori assumptions about the final tensor 
structure of the vertex we formally write (\ref{quardef}) in terms of a basis 
of $138$ Lorentz tensors. This is the number of rank $4$ Lorentz tensors that 
one can build from the three independent external momenta $p$, $q$ and $r$ as 
well as the metric $\eta_{\mu\nu}$. Given the size of this basis, the explicit 
forms given in terms of the labelling we use are recorded in Appendix A and 
denoted by ${\cal P}_{(k) \, \mu \nu \sigma \rho}(p,q,r)$ where $k$ is our 
label with $1$~$\leq$~$k$~$\leq$~$138$. With these then we rewrite 
(\ref{quardef}) as 
\begin{equation}
\left. \Sigma^{abcd}_{\mu \nu \sigma \rho}(p,q,r) 
\right|_{\mbox{\footnotesize{symm}}} ~=~ \sum_{k=1}^{138}
{\cal P}_{(k) \, \mu \nu \sigma \rho}(p,q,r) \, 
\left. \Sigma^{abcd}_{(k)}(p,q,r) \right|_{\mbox{\footnotesize{symm}}} 
\end{equation}
where $\Sigma^{abcd}_{(k)}(p,q,r)$ are the Lorentz scalar but colourful
amplitudes which will be determined. We could choose to write each Lorentz
scalar in terms of colour scalars. However, as this emerges naturally from 
the colour group algebra within the computation for the choice of $SU(\Nc)$ we
have not chosen to proceed that way. Moreover, as the computational method we
will use requires Lorentz scalar Feynman integrals and the colour structure
factors off each integral the projection into Lorentz scalars is more crucial.
For this we follow the earlier method of \cite{25} and first construct the
$138$~$\times$~$138$ matrix ${\cal N}^i_{kl}$ defined by 
\begin{equation}
{\cal N}^i_{kl} ~=~ \left. {\cal P}^i_{(k) \, \mu \nu \sigma \rho}(p,q,r)
{\cal P}^{i ~\, \mu \nu \sigma \rho}_{(l)}(p,q,r) 
\right|_{\mbox{\footnotesize{symm}}} ~.
\end{equation}
If we denote the inverse of ${\cal N}^i_{kl}$ by ${\cal M}^i_{kl}$ then the
latter is the projection tensor we use. In other words 
\begin{equation}
\left. \Sigma^{abcd}_{(k)}(p,q,r) \right|_{\mbox{\footnotesize{symm}}} ~=~
{\cal M}_{kl} \left( {\cal P}^{\mu \nu \sigma \rho}_{(l)}(p,q,r) \left.
\left\langle A^a_\mu(p) A^b_\nu(q) A^c_\sigma(r) A^d_\rho(-p-q-r)
\right\rangle \right ) \right|_{\mbox{\footnotesize{symm}}} 
\label{decomp}
\end{equation}
gives each individual amplitude and there is a sum over $l$.

Once this projection or decomposition (\ref{decomp}) of the Green's function
has been established we need to organize the actual computation. At one loop
there are $24$ Feynman graphs contributing to the quartic function which are
generated using the {\sc Qgraf} package, \cite{34}. Although this is a 
relatively small number given the number of tensors in the projection basis 
there are a large number of individual Feynman integrals within each graph to 
compute. The effect of the projection is to produce scalar integrals which have
at most scalar products in the numerator which depend on the loop momentum and 
the three external momenta. To proceed these are rewritten in terms of the 
possible propagators that arise. The reason for this is that to perform the 
large number of integrals we chose to use the Laporta algorithm, \cite{31}, for
which this is the first step. The algorithm uses integration by parts to 
establish linear relations between integrals within a specific topology. These 
can be solved systematically to produce relations between all the integrals and
a relatively small set of integrals which cannot be reduced to any other
integrals. These are known as master integrals and they have to be evaluated by
direct techniques not involving integration by parts. For the quartic vertex
we have followed this Laporta approach and used its implementation in the
{\sc Reduze} package, \cite{32}. It is a C++ based programme which uses
{\sc GiNaC}, \cite{35}. This allows the user to build a database of requisite
integrals from which one can extract the relations necessary for the particular
Green's function of interest. In order to complete the automatic computation we
use {\sc Form} and its threaded version {\sc Tform}, \cite{36,37}, to handle 
the algebra. The {\sc Reduze} package allows for the integral relations to be 
extracted in {\sc Form} syntax. Indeed {\sc Form} is used to process the
rearrangement of the scalar products in each Feynman graph into {\sc Reduze}
input notation and {\sc Reduze} extracts the correct relations from the
database but in {\sc Form} syntax. After applying the {\sc Reduze} algorithm, 
\cite{32}, there are several basic classes of master integrals to insert from 
direct evaluation. If we regard the number of propagators as defining a class 
then the first set is those integrals with two propagators. This is the simple 
one loop bubble. However, there are two specific integrals which are dependent 
on the value of the square of the momentum flowing through the graph and derive
from how the basic one loop box of four propagators is shrunk. In one instance 
three of the original external legs of the box can be at one external point to 
the bubble and in the other case the external legs of the $2$-propagator bubble
are two pairs of the full box. For the $3$-propagator master, given the 
symmetry of the squared momenta of the external legs there is one master. Here 
two of the legs are at $(-\mu^2)$ while the other is the corresponding 
Mandelstam variable and the value was given in \cite{38,39,40}. Finally, the 
$4$-propagator case is the basic one loop box. The general off-shell expression
was provided in \cite{41}. The explicit forms of the masters in the last two 
classes will be discussed in detail when the results are presented later. As a 
final part of the automatic computation setup description we note that to 
perform the renormalization we follow the method of \cite{42}. This involves 
performing the calculation in terms of bare parameters throughout. Then the 
counterterms are automatically introduced by replacing the bare quantities with
the renormalized counterparts and associated renormalization constant. Those
renormalization constants, such as the wave function and gauge parameter, which
are already known from the renormalization of $2$-point functions are included 
in this redefinition. Hence, one fixes the unknown renormalization constant 
associated with the $4$-point function with the divergences which remain after 
the rescaling.

As we will only be considering $SU(\Nc)$ as the colour rather than a general
Lie group we recall key properties relevant to corrections to the quartic
vertex which are based on \cite{43,18}. As part of the {\sc Qgraf} and
{\sc Form} setup the colour and Lorentz indices are appended to the {\sc Qgraf}
output before the projection and the application of the group theory {\sc Form}
module. First, we recall that the product of two $SU(\Nc)$ group generators, 
$T^a$, can be decomposed as 
\begin{equation}
T^a T^b ~=~ \frac{1}{2\Nc} \delta^{ab} ~+~ \frac{1}{2} d^{abc} T^c ~+~
\frac{i}{2} f^{abc} T^c
\end{equation} 
where $f^{abc}$ are the structure functions and $d^{abc}$ is a totally 
symmetric tensor. The latter vanishes in $SU(2)$. For the other product of
generators, we have 
\begin{equation}
T^a_{IJ} T^a_{KL} ~=~ \frac{1}{2} \left[ \delta_{IL} \delta_{KJ} ~-~ 
\frac{1}{\Nc} \delta_{IJ} \delta_{KL} \right]
\end{equation}
solely for $SU(\Nc)$. To simpify notation we introduce related group tensors
defined by  
\begin{equation}
f_4^{abcd} ~\equiv~ f^{abe} f^{cde} ~~~,~~~ 
d_4^{abcd} ~\equiv~ d^{abe} d^{cde} ~~~,~~~ 
e_4^{abcd} ~\equiv~ d^{abe} f^{cde} ~. 
\end{equation}
So the Jacobi identities, \cite{43}, are readily expressed as
\begin{eqnarray}
e_4^{abcd} ~+~ e_4^{bcad} ~+~ e_4^{cabd} &=& 0 \nonumber \\
f_4^{abcd} ~+~ f_4^{acdb} ~+~ f_4^{adbc} &=& 0 ~.
\label{jac}
\end{eqnarray}
From \cite{43} we note the relation between two products 
\begin{equation}
f_4^{abcd} ~=~ \frac{2}{\Nc} \left[ \delta^{ac} \delta^{bd} ~-~ \delta^{ad}
\delta^{bc} \right] ~+~ d_4^{acbd} ~-~ d_4^{adbc} ~. 
\label{fdec}
\end{equation}
Also since there will be box diagrams with closed quark loops we need the 
decompostion of various traces over the group generators in the fundamental
representation. From \cite{43,18} we use 
\begin{eqnarray}
\mbox{Tr} \left( T^a T^b T^c T^d \right) &=& \frac{1}{16} \left[ 
\frac{4}{\Nc} \left[ \delta^{ab} \delta^{cd} - \delta^{ac} \delta^{bd}
+ \delta^{ad} \delta^{bc} \right]
+ 2 \left[ d_4^{abcd} - d_4^{acbd} + d_4^{adbc} \right]
\right. \nonumber \\
&& \left. ~~~~~
+ 2i \left[ e_4^{abcd} - e_4^{acdb} + e_4^{adbc} \right] \right] ~.
\label{tdec}
\end{eqnarray}
In choosing to rewrite such tensors we are in effect making a choice of how to 
express the group structure of the graphs in the computation. In other words we
eliminate all products of group generators in favour of products which involve
at least one rank three symmetric tensor or the unit matrix. While this is a
suitable choice of colour tensor basis, from the point of view of defining a
MOM type renormalization scheme it is more appropriate to re-express the final
vertex function after computation in terms of the colour tensor structures that
are present in the original vertex in the Lagrangian which is $f_4^{abcd}$ and
one permutation of the indices. The Jacobi identity, (\ref{jac}), can be used
to recover the third tensor. However, due to the symmetry of the quartic vertex
other tensor structures arise at one loop which are not unrelated to the unit
matrices in (\ref{fdec}) and (\ref{tdec}). One could retain these as was the
case in \cite{18} but we have preferred to use the rank four fully symmetric 
tensors introduced in \cite{44}
\begin{equation}
d_F^{abcd} ~=~ \frac{1}{6} \mbox{Tr} \left( T^a T^{(b} T^c T^{d)} \right) ~~,~~
d_A^{abcd} ~=~ \frac{1}{6} \mbox{Tr} \left( T_A^a T_A^{(b} T_A^c T_A^{d)}
\right)
\end{equation}
where the subscript $A$ in the latter denotes the adjoint representation of the
group generator. Hence we have to express $d_4^{abcd}$ in terms of these 
additional colour tensors which is straightforward to do. We have  
\begin{eqnarray}
d_4^{abcd} &=& -~ \frac{1}{3} \left[ f_4^{abcd} - \frac{2}{\Nc}
\left[ \delta^{ac} \delta^{bd} - \delta^{ad} \delta^{bc} \right] \right]
+ \frac{2}{3} \left[ f_4^{acbd} - \frac{2}{\Nc}
\left[ \delta^{ab} \delta^{cd} - \delta^{ad} \delta^{bc} \right] \right]
\nonumber \\
&& +~ 8 \left[ d_F^{acbd} - \frac{1}{12\Nc}
\left[ \delta^{ab} \delta^{cd} + \delta^{ac} \delta^{bd} 
+ \delta^{ad} \delta^{bc} \right] \right]
\nonumber \\
d_4^{adbc} &=& -~ \frac{1}{3} \left[ f_4^{abcd} - \frac{2}{\Nc}
\left[ \delta^{ac} \delta^{bd} - \delta^{ad} \delta^{bc} \right] \right]
- \frac{1}{3} \left[ f_4^{acbd} - \frac{2}{\Nc}
\left[ \delta^{ab} \delta^{cd} - \delta^{ad} \delta^{bc} \right] \right]
\nonumber \\
&& +~ 8 \left[ d_F^{acbd} - \frac{1}{12\Nc}
\left[ \delta^{ab} \delta^{cd} + \delta^{ac} \delta^{bd} 
+ \delta^{ad} \delta^{bc} \right] \right]
\end{eqnarray}
which we use for diagrams involving quark loops and
\begin{eqnarray}
d_4^{abcd} &=& -~ \frac{1}{3} \left[ f_4^{abcd} - \frac{2}{\Nc}
\left[ \delta^{ac} \delta^{bd} - \delta^{ad} \delta^{bc} \right] \right]
+ \frac{2}{3} \left[ f_4^{acbd} - \frac{2}{\Nc}
\left[ \delta^{ab} \delta^{cd} - \delta^{ad} \delta^{bc} \right] \right]
\nonumber \\
&& +~ \frac{4}{\Nc} \left[ d_A^{acbd} - \frac{2}{3}
\left[ \delta^{ab} \delta^{cd} + \delta^{ac} \delta^{bd} 
+ \delta^{ad} \delta^{bc} \right] \right]
\nonumber \\
d_4^{adbc} &=& -~ \frac{1}{3} \left[ f_4^{abcd} - \frac{2}{\Nc}
\left[ \delta^{ac} \delta^{bd} - \delta^{ad} \delta^{bc} \right] \right]
- \frac{1}{3} \left[ f_4^{acbd} - \frac{2}{\Nc}
\left[ \delta^{ab} \delta^{cd} - \delta^{ad} \delta^{bc} \right] \right]
\nonumber \\
&& +~ \frac{4}{\Nc} \left[ d_A^{acbd} - \frac{2}{3}
\left[ \delta^{ab} \delta^{cd} + \delta^{ac} \delta^{bd} 
+ \delta^{ad} \delta^{bc} \right] \right]
\end{eqnarray}
for one loop graphs which do not depend on $\Nf$. Useful identities in
constructing these mappings were, \cite{43},
\begin{equation}
d_4^{abcc} ~=~ 0 ~~~,~~~ 
d_4^{acbc} ~=~ \frac{[\Nc^2-4]}{\Nc} \delta^{ab} ~~~,~~~
d_4^{apbq} d_4^{cdpq} ~=~ \frac{[\Nc^2-12]}{2\Nc} d_4^{abcd} 
\end{equation}
in the present notation. This completes the summary of the computational setup
for the quartic vertex function. 

\sect{Amplitudes.}

We now turn to the task of recording our results and concentrate first on the
amplitudes. Given the size of the final expression for all $138$ channels for
a non-zero gauge parameter $\alpha$ we focus on key parts and relegate the full
results to the attached data file\footnote{Electronic versions of expressions
appearing throughout the article are included in an attached data file.}. (The
Landau gauge corresponds to $\alpha$~$=$~$0$.) To give a flavour of the 
analytic structure which is typical of each channel we record the expressions 
for those which correspond to the Feynman rules for the quartic gluon vertex. 
For the $\MSbar$ scheme we find
\begin{eqnarray}
\Sigma^{abcd}_{(1)}(p,q,r) &=& f_4^{acbd} + f_4^{adbc} \nonumber \\
&& + \left[
\left[
\frac{161}{80}
+ \frac{1}{16} \alpha
- \frac{93}{80} \alpha^2
+ \frac{7}{80} \alpha^3
+ \frac{431}{75} \ln \left( \frac{4}{3} \right)
- \frac{27}{80} \ln \left( \frac{4}{3} \right) \alpha
\right. \right. \nonumber \\
&& \left. \left. ~~~~~
- \frac{253}{200} \ln \left( \frac{4}{3} \right) \alpha^2
+ \frac{9}{400} \ln \left( \frac{4}{3} \right) \alpha^3
- \frac{7219}{38400} \Phi_1 \left( \frac{9}{16}, \frac{9}{16} \right)
\right. \right. \nonumber \\
&& \left. \left. ~~~~~
- \frac{897}{2560} \Phi_1 \left( \frac{9}{16}, \frac{9}{16} \right) \alpha
+ \frac{2709}{12800} \Phi_1 \left( \frac{9}{16}, \frac{9}{16} \right) \alpha^2
\right. \right. \nonumber \\
&& \left. \left. ~~~~~
- \frac{351}{12800} \Phi_1 \left( \frac{9}{16}, \frac{9}{16} \right) \alpha^3
- \frac{85}{192} \Phi_1 \left( \frac{3}{4}, \frac{3}{4} \right)
- \frac{5}{32} \Phi_1 \left( \frac{3}{4}, \frac{3}{4} \right) \alpha
\right. \right. \nonumber \\
&& \left. \left. ~~~~~
+ \frac{9}{64} \Phi_1 \left( \frac{3}{4}, \frac{3}{4} \right) \alpha^2
\right] f_4^{acbd} \Nc 
\right. \nonumber \\
&& \left. ~~~
+ \left[
\frac{2}{3}
- \frac{7}{6} \ln \left( \frac{4}{3} \right)
+ \frac{1}{24} \Phi_1 \left( \frac{9}{16}, \frac{9}{16} \right)
- \frac{7}{24} \Phi_1 \left( \frac{3}{4}, \frac{3}{4} \right)
\right] f_4^{acbd} \Nf
\right. \nonumber \\
&& \left. ~~~
+ \left[
 \frac{161}{80}
+ \frac{1}{16} \alpha
- \frac{93}{80} \alpha^2
+ \frac{7}{80} \alpha^3
+ \frac{431}{75} \ln \left( \frac{4}{3} \right)
- \frac{27}{80} \ln \left( \frac{4}{3} \right) \alpha
\right. \right. \nonumber \\
&& \left. \left. ~~~~~~~~
- \frac{253}{200} \ln \left( \frac{4}{3} \right) \alpha^2
+ \frac{9}{400} \ln \left( \frac{4}{3} \right) \alpha^3
- \frac{7219}{38400} \Phi_1 \left( \frac{9}{16}, \frac{9}{16} \right)
\right. \right. \nonumber \\
&& \left. \left. ~~~~~~~~
- \frac{897}{2560} \Phi_1 \left( \frac{9}{16}, \frac{9}{16} \right) \alpha
+ \frac{2709}{12800} \Phi_1 \left( \frac{9}{16}, \frac{9}{16} \right) \alpha^2
\right. \right. \nonumber \\
&& \left. \left. ~~~~~~~~
- \frac{351}{12800} \Phi_1 \left( \frac{9}{16}, \frac{9}{16} \right) \alpha^3
- \frac{85}{192} \Phi_1 \left( \frac{3}{4}, \frac{3}{4} \right)
- \frac{5}{32} \Phi_1 \left( \frac{3}{4}, \frac{3}{4} \right) \alpha
\right. \right. \nonumber \\
&& \left. \left. ~~~~~~~~
+ \frac{9}{64} \Phi_1 \left( \frac{3}{4}, \frac{3}{4} \right) \alpha^2
\right] f_4^{adbc} \Nc
\right. \nonumber \\
&& \left. ~~~
+ \left[
\frac{2}{3}
- \frac{7}{6} \ln \left( \frac{4}{3} \right)
+ \frac{1}{24} \Phi_1 \left( \frac{9}{16}, \frac{9}{16} \right)
- \frac{7}{24} \Phi_1 \left( \frac{3}{4}, \frac{3}{4} \right)
\right] f_4^{adbc} \Nf 
\right. \nonumber \\
&& \left. ~~~
+ \left[
\frac{3577}{1920}
- \frac{123}{160} \alpha
+ \frac{21}{64} \alpha^2
- \frac{15}{32} \alpha^3
+ \frac{243}{640} \alpha^4
+ \frac{28113}{3200} \ln \left( \frac{4}{3} \right)
\right. \right. \nonumber \\
&& \left. \left. ~~~~~~~~
- \frac{903}{400} \ln \left( \frac{4}{3} \right) \alpha
- \frac{867}{320} \ln \left( \frac{4}{3} \right) \alpha^2
- \frac{27}{20} \ln \left( \frac{4}{3} \right) \alpha^3
+ \frac{1701}{3200} \ln \left( \frac{4}{3} \right) \alpha^4
\right. \right. \nonumber \\
&& \left. \left. ~~~~~~~~
+ \frac{115493}{102400} \Phi_1 \left( \frac{9}{16}, \frac{9}{16} \right)
- \frac{5841}{25600} \Phi_1 \left( \frac{9}{16}, \frac{9}{16} \right) \alpha
\right. \right. \nonumber \\
&& \left. \left. ~~~~~~~~
+ \frac{16443}{10240} \Phi_1 \left( \frac{9}{16}, \frac{9}{16} \right) \alpha^2
- \frac{1701}{5120} \Phi_1 \left( \frac{9}{16}, \frac{9}{16} \right) \alpha^3
\right. \right. \nonumber \\
&& \left. \left. ~~~~~~~~
- \frac{17739}{102400} \Phi_1 \left( \frac{9}{16}, \frac{9}{16} \right) \alpha^4
- \frac{145}{128} \Phi_1 \left( \frac{3}{4}, \frac{3}{4} \right)
+ \frac{147}{128} \Phi_1 \left( \frac{3}{4}, \frac{3}{4} \right) \alpha
\right. \right. \nonumber \\
&& \left. \left. ~~~~~~~~
- \frac{243}{128} \Phi_1 \left( \frac{3}{4}, \frac{3}{4} \right) \alpha^2
+ \frac{81}{128} \Phi_1 \left( \frac{3}{4}, \frac{3}{4} \right) \alpha^3
\right] d_A^{abcd}
\right. \nonumber \\
&& \left. ~~~
+ \left[
-~ \frac{8}{3}
- 6 \ln \left( \frac{4}{3} \right)
- \Phi_1 \left( \frac{9}{16}, \frac{9}{16} \right)
+ \frac{5}{2} \Phi_1 \left( \frac{3}{4}, \frac{3}{4} \right)
\right] d_F^{abcd} \Nf
\right] a \nonumber \\
&& +~ O(a^2)
\label{msamp}
\end{eqnarray}
where we have used the Jacobi identity at one loop to recover the symmetry 
structure of the quartic vertex Feynman rule. For the other two channels given 
the similarity of the expressions we note that they satisfy
\begin{equation}
\Sigma^{abcd}_{(2)}(p,q,r) ~=~ \Sigma^{acbd}_{(1)}(p,q,r) ~~,~~
\Sigma^{abcd}_{(3)}(p,q,r) ~=~ \Sigma^{adcb}_{(1)}(p,q,r) 
\end{equation}
which is a useful check on the computation. Other checks include the fact that
the one loop Green's function is finite when the $\MSbar$ gluon wave function 
and coupling constant renormalization constants are included. The two specific
values of the function $\Phi_1(x,y)$ appearing in (\ref{msamp}) derive from the
two master integrals referred to already. The general expression for
$\Phi_1(x,y)$ includes the usual dilogarithm function $\mbox{Li}_2(z)$ via,
\cite{38,39}, 
\begin{equation}
\Phi_1(x,y) ~=~ \frac{1}{\lambda} \left[ 2 \mbox{Li}_2(-\rho x)
+ 2 \mbox{Li}_2(-\rho y)
+ \ln \left( \frac{y}{x} \right)
\ln \left( \frac{(1+\rho y)}{(1+\rho x)} \right)
+ \ln(\rho x) \ln(\rho y) + \frac{\pi^2}{3} \right] 
\label{phi1def}
\end{equation}
where
\begin{equation}
\lambda(x,y) ~=~ \sqrt{\Delta_G} ~~~,~~~
\rho(x,y) ~=~ \frac{2}{[1-x-y+\lambda(x,y)]}
\end{equation}
and 
\begin{equation}
\Delta_G(x,y) ~=~ x^2 ~-~ 2 x y ~+~ y^2 ~-~ 2 x ~-~ 2 y ~+~ 1
\end{equation}
is the Gram determinant. The appearance of $\Phi_1(x,y)$ at two different but 
symmetric arguments arise from two masters in the Laporta sense. One is for the
$3$-point function where the squares of the external momenta are $p^2$, $q^2$ 
and $s$ and is $\Phi_1 \left( \frac{3}{4},\frac{3}{4} \right)$. The other is 
for the pure symmetric scalar box which was computed in \cite{41} corresponding
to $\Phi_1 \left( \frac{9}{16},\frac{9}{16} \right)$. When one evaluates these 
functions from (\ref{phi1def}) the dilogarithms involve the Clausen function,
$\mbox{Cl}_2(\theta)$, since the argument of the dilogarithm is complex. Though
the expression for each is ultimately real. In particular the argument 
of each dilogarithm function is $\frac{1}{3}(1+2\sqrt{2}i)$ for the triangle 
master and $\frac{1}{9}(1+4\sqrt{5}i)$ for the box master. Using the more
symmetric definition of $\Phi_1(x,y)$ given in \cite{45}, our two basic master
values can be written as
\begin{eqnarray}
\Phi_1 \left( \frac{3}{4},\frac{3}{4} \right) &=& \sqrt{2} \left[ 
2 \mbox{Cl}_2 \left( 2 \cos^{-1} \left( \frac{1}{\sqrt{3}} \right) \right) 
+ \mbox{Cl}_2 \left( 2 \cos^{-1} \left( \frac{1}{3} \right) \right) \right] 
\nonumber \\ 
\Phi_1 \left( \frac{9}{16},\frac{9}{16} \right) &=& \frac{4}{\sqrt{5}} \left[ 
2 \mbox{Cl}_2 \left( 2 \cos^{-1} \left( \frac{2}{3} \right) \right) 
+ \mbox{Cl}_2 \left( 2 \cos^{-1} \left( \frac{1}{9} \right) \right) \right] ~. 
\end{eqnarray}
Though, for compactness in presenting results we will use the shorthand 
notation $\Phi_1 \left( \frac{3}{4},\frac{3}{4} \right)$ and 
$\Phi_1 \left( \frac{9}{16},\frac{9}{16} \right)$ throughout. 

The other main check on our one loop expression is to compare with the early
work of \cite{18}. In that article the quartic vertex was used to construct
a renormalization scheme motivated by ideas of Weinberg, \cite{46}. More
specifically a coupling constant renormalization, denoted there by $Z_5$, for 
this Weinberg scheme was recorded for an arbitrary $\alpha$ and we have been 
able to virtually reproduce it. However, in order to do so we have had to 
convert to the same colour tensor basis as \cite{18} which involved $d^{abc}$ 
and the unit matrix rather than our $\{f_4^{abcd},d_F^{abcd},d_A^{abcd}\}$ 
basis. This is straightforward to do. Also we have had to map our two main 
master functions to those present in \cite{18} which are 
$R \left( \frac{2}{3} \right)$ and $K \left( \frac{2}{3} \right)$ in the 
notation of \cite{18}. The relations between these and those which appear in 
our computation are
\begin{equation}
R \left( \frac{2}{3} \right) ~=~ \frac{3}{4} \Phi_1 \left( \frac{3}{4}, 
\frac{3}{4} \right) ~~~,~~~ 
K \left( \frac{2}{3} \right) ~=~ R \left( \frac{2}{3} \right) ~-~
\frac{15}{32} \Phi_1 \left( \frac{9}{16}, \frac{9}{16} \right) ~. 
\label{ptmap}
\end{equation}
However in \cite{18} the values of the functions were only expressed 
numerically and not in analytic form involving $\mbox{Li}_2(z)$. Using 
(\ref{phi1def}) we have checked that the numerical values of (\ref{ptmap}) are 
in agreement with the values given in \cite{18}. For completeness and for a 
numerical evaluation we note that 
\begin{equation}
\Phi_1 \left( \frac{3}{4}, \frac{3}{4} \right) ~=~ 2.832045 ~~~,~~~
\Phi_1 \left( \frac{9}{16}, \frac{9}{16} \right) ~=~ 3.403614 ~. 
\label{ptmapval}
\end{equation}
In extracting the same renormalization constant $Z_5$ as \cite{18} from our 
analysis we find agreement with all the terms given in equation (3.10) of
\cite{18}, including that noted in the erratum associated with \cite{18}, 
except for two coefficients in the term linear in what corresponds to our 
$\alpha$. Given that we get agreement with the other $18$ coefficients which 
have various powers of $\alpha$ we assume that there is a minor discrepancy in 
\cite{18}. For the interested reader we believe the correct coefficients are 
obtained if $\frac{315}{20}$ and $\frac{163}{100}$ given in \cite{18} are 
replaced by $\frac{84}{5}$ and $\frac{22}{25}$ respectively. Although we are 
not in a position to comment on the effect these changes would have on the 
subsequent analysis performed in \cite{18}, we note that the numerical value of
this term linear in $\alpha$ increases by about $10\%$ primarily due to the 
large drop in the second value. Aside from these two terms and in light of the 
exact agreement with the other terms in $Z_5$ of \cite{18}, we believe we have 
the correct form for the one loop amplitudes. 

In order to gauge the structure of the full one loop vertex function we have
evaluated it numerically in the Landau gauge and find
\begin{eqnarray}
\!\! \left. \Sigma^{abcd}(p,q,r) \right|_{\alpha=0} &=&
f_4^{abcd} \left[ {\cal P}_{(2)}
- {\cal P}_{(3)} \right]
+ f_4^{acbd} \left[ {\cal P}_{(1)}
- {\cal P}_{(3)} \right]
+ f_4^{adbc} \left[ {\cal P}_{(1)}
- {\cal P}_{(2)} \right]
\nonumber \\
&& 
+ \left[ \left[ 
5.021026 {\cal P}_{(1)}
+ 5.021026 {\cal P}_{(2)}
+ 5.021026 {\cal P}_{(3)}
+ 4.028121 {\cal P}_{(4)}
\right. \right. \nonumber \\
&& \left. \left. ~~~~
+ 6.174705 {\cal P}_{(5)}
- 1.359046 {\cal P}_{(6)}
+ 6.174705 {\cal P}_{(7)}
+ 4.028121 {\cal P}_{(8)}
\right. \right. \nonumber \\
&& \left. \left. ~~~~
- 1.359046 {\cal P}_{(9)}
+ 3.872502 {\cal P}_{(10)}
+ 3.872502 {\cal P}_{(11)}
+ 2.513456 {\cal P}_{(12)}
\right. \right. \nonumber \\
&& \left. \left. ~~~~
+ 4.028121 {\cal P}_{(13)}
- 1.359046 {\cal P}_{(14)}
+ 6.174705 {\cal P}_{(15)}
+ 3.872502 {\cal P}_{(16)}
\right. \right. \nonumber \\
&& \left. \left. ~~~~
+ 2.513456 {\cal P}_{(17)}
+ 3.872502 {\cal P}_{(18)}
+ 6.174705 {\cal P}_{(19)}
- 1.359046 {\cal P}_{(20)}
\right. \right. \nonumber \\
&& \left. \left. ~~~~
+ 4.028121 {\cal P}_{(21)}
- 4.293168 {\cal P}_{(22)}
- 2.146584 {\cal P}_{(23)}
- 2.146584 {\cal P}_{(24)}
\right. \right. \nonumber \\
&& \left. \left. ~~~~
- 2.146584 {\cal P}_{(25)}
+ 4.028121 {\cal P}_{(26)}
+ 5.387167 {\cal P}_{(27)}
- 2.146584 {\cal P}_{(28)}
\right. \right. \nonumber \\
&& \left. \left. ~~~~
+ 0.155618 {\cal P}_{(29)}
+ 4.028121 {\cal P}_{(30)}
+ 2.513456 {\cal P}_{(31)}
+ 3.872502 {\cal P}_{(32)}
\right. \right. \nonumber \\
&& \left. \left. ~~~~
+ 3.872502 {\cal P}_{(33)}
- 1.359046 {\cal P}_{(34)}
+ 4.028121 {\cal P}_{(35)}
+ 6.174705 {\cal P}_{(36)}
\right. \right. \nonumber \\
&& \left. \left. ~~~~
- 1.359046 {\cal P}_{(37)}
+ 6.174705 {\cal P}_{(38)}
+ 4.028121 {\cal P}_{(39)}
+ 4.028121 {\cal P}_{(40)}
\right. \right. \nonumber \\
&& \left. \left. ~~~~
- 2.146584 {\cal P}_{(41)}
+ 0.155618 {\cal P}_{(42)}
- 2.146584 {\cal P}_{(43)}
- 4.293168 {\cal P}_{(44)}
\right. \right. \nonumber \\
&& \left. \left. ~~~~
- 2.146584 {\cal P}_{(45)}
+ 5.387167 {\cal P}_{(46)}
- 2.146584 {\cal P}_{(47)}
+ 4.028121 {\cal P}_{(48)}
\right. \right. \nonumber \\
&& \left. \left. ~~~~
+ 4.028121 {\cal P}_{(49)}
+ 0.155618 {\cal P}_{(50)}
- 2.146584 {\cal P}_{(51)}
+ 5.387167 {\cal P}_{(52)}
\right. \right. \nonumber \\
&& \left. \left. ~~~~
+ 4.028121 {\cal P}_{(53)}
- 2.146584 {\cal P}_{(54)}
- 2.146584 {\cal P}_{(55)}
- 2.146584 {\cal P}_{(56)}
\right. \right. \nonumber \\
&& \left. \left. ~~~~
- 4.293168 {\cal P}_{(57)}
+ 0.397845 {\cal P}_{(58)}
+ 0.397845 {\cal P}_{(59)}
+ 0.397845 {\cal P}_{(60)}
\right. \right. \nonumber \\
&& \left. \left. ~~~~
+ 2.493711 {\cal P}_{(61)}
+ 0.198922 {\cal P}_{(62)}
+ 0.198922 {\cal P}_{(63)}
- 2.095866 {\cal P}_{(64)}
\right. \right. \nonumber \\
&& \left. \left. ~~~~
+ 2.493711 {\cal P}_{(65)}
+ 0.198922 {\cal P}_{(66)}
+ 0.198922 {\cal P}_{(67)}
- 2.095866 {\cal P}_{(68)}
\right. \right. \nonumber \\
&& \left. \left. ~~~~
+ 2.493711 {\cal P}_{(69)}
+ 0.198922 {\cal P}_{(70)}
- 2.095866 {\cal P}_{(71)}
+ 0.198922 {\cal P}_{(72)}
\right. \right. \nonumber \\
&& \left. \left. ~~~~
+ 2.493711 {\cal P}_{(73)}
- 2.095866 {\cal P}_{(74)}
+ 0.198922 {\cal P}_{(75)}
+ 0.198922 {\cal P}_{(76)}
\right. \right. \nonumber \\
&& \left. \left. ~~~~
+ 2.493711 {\cal P}_{(77)}
- 2.095866 {\cal P}_{(78)}
+ 0.198922 {\cal P}_{(79)}
+ 0.198922 {\cal P}_{(80)}
\right. \right. \nonumber \\
&& \left. \left. ~~~~
+ 2.493711 {\cal P}_{(81)}
+ 0.198922 {\cal P}_{(82)}
- 2.095866 {\cal P}_{(83)}
+ 0.198922 {\cal P}_{(84)}
\right. \right. \nonumber \\
&& \left. \left. ~~~~
+ 0.198922 {\cal P}_{(85)}
+ 2.493711 {\cal P}_{(86)}
- 2.095866 {\cal P}_{(87)}
+ 2.493711 {\cal P}_{(88)}
\right. \right. \nonumber \\
&& \left. \left. ~~~~
- 2.095866 {\cal P}_{(89)}
+ 0.198922 {\cal P}_{(90)}
+ 2.493711 {\cal P}_{(91)}
+ 0.198922 {\cal P}_{(92)}
\right. \right. \nonumber \\
&& \left. \left. ~~~~
- 2.095866 {\cal P}_{(93)}
+ 2.493711 {\cal P}_{(94)}
+ 0.198922 {\cal P}_{(95)}
- 2.095866 {\cal P}_{(96)}
\right. \right. \nonumber \\
&& \left. \left. ~~~~
+ 2.493711 {\cal P}_{(97)}
- 2.095866 {\cal P}_{(98)}
+ 0.198922 {\cal P}_{(99)}
+ 0.198922 {\cal P}_{(100)}
\right. \right. \nonumber \\
&& \left. \left. ~~~~
+ 2.493711 {\cal P}_{(101)}
- 2.095866 {\cal P}_{(102)}
+ 3.917581 {\cal P}_{(103)}
+ 2.663495 {\cal P}_{(104)}
\right. \right. \nonumber \\
&& \left. \left. ~~~~
- 0.956684 {\cal P}_{(105)}
+ 0.681225 {\cal P}_{(106)}
+ 0.368706 {\cal P}_{(107)}
+ 1.622792 {\cal P}_{(108)}
\right. \right. \nonumber \\
&& \left. \left. ~~~~
+ 2.663495 {\cal P}_{(109)}
+ 3.917581 {\cal P}_{(110)}
- 0.956684 {\cal P}_{(111)}
- 2.210770 {\cal P}_{(112)}
\right. \right. \nonumber \\
&& \left. \left. ~~~~
+ 1.622792 {\cal P}_{(113)}
+ 0.368706 {\cal P}_{(114)}
+ 2.663495 {\cal P}_{(115)}
- 0.956684 {\cal P}_{(116)}
\right. \right. \nonumber \\
&& \left. \left. ~~~~
+ 3.917581 {\cal P}_{(117)}
+ 1.622792 {\cal P}_{(118)}
- 2.210770 {\cal P}_{(119)}
+ 1.622792 {\cal P}_{(120)}
\right. \right. \nonumber \\
&& \left. \left. ~~~~
+ 3.917581 {\cal P}_{(121)}
- 0.956684 {\cal P}_{(122)}
+ 2.663495 {\cal P}_{(123)}
+ 0.368706 {\cal P}_{(124)}
\right. \right. \nonumber \\
&& \left. \left. ~~~~
+ 0.681225 {\cal P}_{(125)}
+ 0.368706 {\cal P}_{(126)}
- 0.956684 {\cal P}_{(127)}
+ 3.917581 {\cal P}_{(128)}
\right. \right. \nonumber \\
&& \left. \left. ~~~~
+ 2.663495 {\cal P}_{(129)}
+ 1.622792 {\cal P}_{(130)}
+ 0.368706 {\cal P}_{(131)}
+ 0.681225 {\cal P}_{(132)}
\right. \right. \nonumber \\
&& \left. \left. ~~~~
- 0.956684 {\cal P}_{(133)}
+ 2.663495 {\cal P}_{(134)}
+ 3.917581 {\cal P}_{(135)}
+ 0.368706 {\cal P}_{(136)}
\right. \right. \nonumber \\
&& \left. \left. ~~~~
+ 1.622792 {\cal P}_{(137)}
- 2.210770 {\cal P}_{(138)}
\right] d_A^{abcd} 
\right. \nonumber \\
&& \left. 
+ \left[ 
- 0.716260 {\cal P}_{(1)}
- 0.716260 {\cal P}_{(2)}
- 0.716260 {\cal P}_{(3)}
- 2.382957 {\cal P}_{(4)}
\right. \right. \nonumber \\
&& \left. \left. ~~~~
- 4.268597 {\cal P}_{(5)}
+ 4.502776 {\cal P}_{(6)}
- 4.268597 {\cal P}_{(7)}
- 2.382957 {\cal P}_{(8)}
\right. \right. \nonumber \\
&& \left. \left. ~~~~
+ 4.502776 {\cal P}_{(9)}
- 2.217185 {\cal P}_{(10)}
- 2.217185 {\cal P}_{(11)}
+ 2.285591 {\cal P}_{(12)}
\right. \right. \nonumber \\
&& \left. \left. ~~~~
- 2.382957 {\cal P}_{(13)}
+ 4.502776 {\cal P}_{(14)}
- 4.268597 {\cal P}_{(15)}
- 2.217185 {\cal P}_{(16)}
\right. \right. \nonumber \\
&& \left. \left. ~~~~
+ 2.285591 {\cal P}_{(17)}
- 2.217185 {\cal P}_{(18)}
- 4.268597 {\cal P}_{(19)}
+ 4.502776 {\cal P}_{(20)}
\right. \right. \nonumber \\
&& \left. \left. ~~~~
- 2.382957 {\cal P}_{(21)}
+ 3.771280 {\cal P}_{(22)}
+ 1.885640 {\cal P}_{(23)}
+ 1.885640 {\cal P}_{(24)}
\right. \right. \nonumber \\
&& \left. \left. ~~~~
+ 1.885640 {\cal P}_{(25)}
- 2.382957 {\cal P}_{(26)}
- 6.885733 {\cal P}_{(27)}
+ 1.885640 {\cal P}_{(28)}
\right. \right. \nonumber \\
&& \left. \left. ~~~~
- 0.165772 {\cal P}_{(29)}
- 2.382957 {\cal P}_{(30)}
+ 2.285591 {\cal P}_{(31)}
- 2.217185 {\cal P}_{(32)}
\right. \right. \nonumber \\
&& \left. \left. ~~~~
- 2.217185 {\cal P}_{(33)}
+ 4.502776 {\cal P}_{(34)}
- 2.382957 {\cal P}_{(35)}
- 4.268597 {\cal P}_{(36)}
\right. \right. \nonumber \\
&& \left. \left. ~~~~
+ 4.502776 {\cal P}_{(37)}
- 4.268597 {\cal P}_{(38)}
- 2.382957 {\cal P}_{(39)}
- 2.382957 {\cal P}_{(40)}
\right. \right. \nonumber \\
&& \left. \left. ~~~~
+ 1.885640 {\cal P}_{(41)}
- 0.165772 {\cal P}_{(42)}
+ 1.885640 {\cal P}_{(43)}
+ 3.771280 {\cal P}_{(44)}
\right. \right. \nonumber \\
&& \left. \left. ~~~~
+ 1.885640 {\cal P}_{(45)}
- 6.885733 {\cal P}_{(46)}
+ 1.885640 {\cal P}_{(47)}
- 2.382957 {\cal P}_{(48)}
\right. \right. \nonumber \\
&& \left. \left. ~~~~
- 2.382957 {\cal P}_{(49)}
- 0.165772 {\cal P}_{(50)}
+ 1.885640 {\cal P}_{(51)}
- 6.885733 {\cal P}_{(52)}
\right. \right. \nonumber \\
&& \left. \left. ~~~~
- 2.382957 {\cal P}_{(53)}
+ 1.885640 {\cal P}_{(54)}
+ 1.885640 {\cal P}_{(55)}
+ 1.885640 {\cal P}_{(56)}
\right. \right. \nonumber \\
&& \left. \left. ~~~~
+ 3.771280 {\cal P}_{(57)}
- 1.307807 {\cal P}_{(58)}
- 1.307807 {\cal P}_{(59)}
- 1.307807 {\cal P}_{(60)}
\right. \right. \nonumber \\
&& \left. \left. ~~~~
- 0.838049 {\cal P}_{(61)}
- 0.653903 {\cal P}_{(62)}
- 0.653903 {\cal P}_{(63)}
- 0.469758 {\cal P}_{(64)}
\right. \right. \nonumber \\
&& \left. \left. ~~~~
- 0.838049 {\cal P}_{(65)}
- 0.653903 {\cal P}_{(66)}
- 0.653903 {\cal P}_{(67)}
- 0.469758 {\cal P}_{(68)}
\right. \right. \nonumber \\
&& \left. \left. ~~~~
- 0.838049 {\cal P}_{(69)}
- 0.653903 {\cal P}_{(70)}
- 0.469758 {\cal P}_{(71)}
- 0.653903 {\cal P}_{(72)}
\right. \right. \nonumber \\
&& \left. \left. ~~~~
- 0.838049 {\cal P}_{(73)}
- 0.469758 {\cal P}_{(74)}
- 0.653903 {\cal P}_{(75)}
- 0.653903 {\cal P}_{(76)}
\right. \right. \nonumber \\
&& \left. \left. ~~~~
- 0.838049 {\cal P}_{(77)}
- 0.469758 {\cal P}_{(78)}
- 0.653903 {\cal P}_{(79)}
- 0.653903 {\cal P}_{(80)}
\right. \right. \nonumber \\
&& \left. \left. ~~~~
- 0.838049 {\cal P}_{(81)}
- 0.653903 {\cal P}_{(82)}
- 0.469758 {\cal P}_{(83)}
- 0.653903 {\cal P}_{(84)}
\right. \right. \nonumber \\
&& \left. \left. ~~~~
- 0.653903 {\cal P}_{(85)}
- 0.838049 {\cal P}_{(86)}
- 0.469758 {\cal P}_{(87)}
- 0.838049 {\cal P}_{(88)}
\right. \right. \nonumber \\
&& \left. \left. ~~~~
- 0.469758 {\cal P}_{(89)}
- 0.653903 {\cal P}_{(90)}
- 0.838049 {\cal P}_{(91)}
- 0.653903 {\cal P}_{(92)}
\right. \right. \nonumber \\
&& \left. \left. ~~~~
- 0.469758 {\cal P}_{(93)}
- 0.838049 {\cal P}_{(94)}
- 0.653903 {\cal P}_{(95)}
- 0.469758 {\cal P}_{(96)}
\right. \right. \nonumber \\
&& \left. \left. ~~~~
- 0.838049 {\cal P}_{(97)}
- 0.469758 {\cal P}_{(98)}
- 0.653903 {\cal P}_{(99)}
- 0.653903 {\cal P}_{(100)}
\right. \right. \nonumber \\
&& \left. \left. ~~~~
- 0.838049 {\cal P}_{(101)}
- 0.469758 {\cal P}_{(102)}
+ 5.472775 {\cal P}_{(103)}
- 0.497317 {\cal P}_{(104)}
\right. \right. \nonumber \\
&& \left. \left. ~~~~
- 2.746924 {\cal P}_{(105)}
- 0.497317 {\cal P}_{(106)}
- 0.313172 {\cal P}_{(107)}
+ 5.656920 {\cal P}_{(108)}
\right. \right. \nonumber \\
&& \left. \left. ~~~~
- 0.497317 {\cal P}_{(109)}
+ 5.472775 {\cal P}_{(110)}
- 2.746924 {\cal P}_{(111)}
- 8.717016 {\cal P}_{(112)}
\right. \right. \nonumber \\
&& \left. \left. ~~~~
+ 5.656920 {\cal P}_{(113)}
- 0.313172 {\cal P}_{(114)}
- 0.497317 {\cal P}_{(115)}
- 2.746924 {\cal P}_{(116)}
\right. \right. \nonumber \\
&& \left. \left. ~~~~
+ 5.472775 {\cal P}_{(117)}
+ 5.656920 {\cal P}_{(118)}
- 8.717016 {\cal P}_{(119)}
+ 5.656920 {\cal P}_{(120)}
\right. \right. \nonumber \\
&& \left. \left. ~~~~
+ 5.472775 {\cal P}_{(121)}
- 2.746924 {\cal P}_{(122)}
- 0.497317 {\cal P}_{(123)}
- 0.313172 {\cal P}_{(124)}
\right. \right. \nonumber \\
&& \left. \left. ~~~~
- 0.497317 {\cal P}_{(125)}
- 0.313172 {\cal P}_{(126)}
- 2.746924 {\cal P}_{(127)}
+ 5.472775 {\cal P}_{(128)}
\right. \right. \nonumber \\
&& \left. \left. ~~~~
- 0.497317 {\cal P}_{(129)}
+ 5.656920 {\cal P}_{(130)}
- 0.313172 {\cal P}_{(131)}
- 0.497317 {\cal P}_{(132)}
\right. \right. \nonumber \\
&& \left. \left. ~~~~
- 2.746924 {\cal P}_{(133)}
- 0.497317 {\cal P}_{(134)}
+ 5.472775 {\cal P}_{(135)}
- 0.313172 {\cal P}_{(136)}
\right. \right. \nonumber \\
&& \left. \left. ~~~~
+ 5.656920 {\cal P}_{(137)}
- 8.717016 {\cal P}_{(138)}
\right] d_F^{abcd} \Nf 
\right. \nonumber \\
&& \left. 
+ \left[ 
- 0.353158 {\cal P}_{(2)}
+ 0.353158 {\cal P}_{(3)}
+ 0.396705 {\cal P}_{(4)}
- 0.086664 {\cal P}_{(5)}
\right. \right. \nonumber \\
&& \left. \left. ~~~~
+ 0.374322 {\cal P}_{(6)}
+ 0.861923 {\cal P}_{(7)}
- 0.009757 {\cal P}_{(8)}
- 0.032141 {\cal P}_{(9)}
\right. \right. \nonumber \\
&& \left. \left. ~~~~
+ 0.545727 {\cal P}_{(10)}
+ 0.139265 {\cal P}_{(11)}
+ 0.513587 {\cal P}_{(12)}
- 0.386947 {\cal P}_{(13)}
\right. \right. \nonumber \\
&& \left. \left. ~~~~
- 0.342181 {\cal P}_{(14)}
- 0.775260 {\cal P}_{(15)}
- 0.684993 {\cal P}_{(16)}
- 1.027174 {\cal P}_{(17)}
\right. \right. \nonumber \\
&& \left. \left. ~~~~
- 0.684993 {\cal P}_{(18)}
- 0.775259 {\cal P}_{(19)}
- 0.342181 {\cal P}_{(20)}
- 0.386947 {\cal P}_{(21)}
\right. \right. \nonumber \\
&& \left. \left. ~~~~
- 0.388312 {\cal P}_{(22)}
+ 0.076906 {\cal P}_{(23)}
+ 0.483368 {\cal P}_{(24)}
- 0.871680 {\cal P}_{(25)}
\right. \right. \nonumber \\
&& \left. \left. ~~~~
- 0.009757 {\cal P}_{(26)}
+ 0.022383 {\cal P}_{(27)}
- 0.465218 {\cal P}_{(28)}
- 0.149022 {\cal P}_{(29)}
\right. \right. \nonumber \\
&& \left. \left. ~~~~
+ 0.396704 {\cal P}_{(30)}
+ 0.513586 {\cal P}_{(31)}
+ 0.139265 {\cal P}_{(32)}
+ 0.545727 {\cal P}_{(33)}
\right. \right. \nonumber \\
&& \left. \left. ~~~~
- 0.032140 {\cal P}_{(34)}
- 0.009757 {\cal P}_{(35)}
+ 0.861923 {\cal P}_{(36)}
+ 0.374321 {\cal P}_{(37)}
\right. \right. \nonumber \\
&& \left. \left. ~~~~
- 0.086664 {\cal P}_{(38)}
+ 0.396704 {\cal P}_{(39)}
- 0.386947 {\cal P}_{(40)}
+ 0.388311 {\cal P}_{(41)}
\right. \right. \nonumber \\
&& \left. \left. ~~~~
+ 0.298045 {\cal P}_{(42)}
+ 0.388311 {\cal P}_{(43)}
+ 0.776623 {\cal P}_{(44)}
+ 0.388311 {\cal P}_{(45)}
\right. \right. \nonumber \\
&& \left. \left. ~~~~
- 0.044766 {\cal P}_{(46)}
+ 0.388311 {\cal P}_{(47)}
- 0.386947 {\cal P}_{(48)}
+ 0.396704 {\cal P}_{(49)}
\right. \right. \nonumber \\
&& \left. \left. ~~~~
- 0.149022 {\cal P}_{(50)}
- 0.465218 {\cal P}_{(51)}
+ 0.022383 {\cal P}_{(52)}
- 0.009757 {\cal P}_{(53)}
\right. \right. \nonumber \\
&& \left. \left. ~~~~
- 0.871680 {\cal P}_{(54)}
+ 0.483368 {\cal P}_{(55)}
+ 0.076906 {\cal P}_{(56)}
- 0.388311 {\cal P}_{(57)}
\right. \right. \nonumber \\
&& \left. \left. ~~~~
+ 0.553863 {\cal P}_{(58)}
- 1.107726 {\cal P}_{(59)}
+ 0.553863 {\cal P}_{(60)}
+ 0.596522 {\cal P}_{(61)}
\right. \right. \nonumber \\
&& \left. \left. ~~~~
+ 0.844398 {\cal P}_{(62)}
+ 0.168365 {\cal P}_{(63)}
- 0.048819 {\cal P}_{(64)}
+ 0.602682 {\cal P}_{(65)}
\right. \right. \nonumber \\
&& \left. \left. ~~~~
+ 0.385497 {\cal P}_{(66)}
- 0.290535 {\cal P}_{(67)}
- 0.042659 {\cal P}_{(68)}
- 1.199205 {\cal P}_{(69)}
\right. \right. \nonumber \\
&& \left. \left. ~~~~
- 0.553863 {\cal P}_{(70)}
+ 0.091478 {\cal P}_{(71)}
- 0.553863 {\cal P}_{(72)}
+ 0.596522 {\cal P}_{(73)}
\right. \right. \nonumber \\
&& \left. \left. ~~~~
- 0.048819 {\cal P}_{(74)}
+ 0.168365 {\cal P}_{(75)}
+ 0.844398 {\cal P}_{(76)}
+ 0.602682 {\cal P}_{(77)}
\right. \right. \nonumber \\
&& \left. \left. ~~~~
- 0.042659 {\cal P}_{(78)}
- 0.290535 {\cal P}_{(79)}
+ 0.385497 {\cal P}_{(80)}
- 1.199205 {\cal P}_{(81)}
\right. \right. \nonumber \\
&& \left. \left. ~~~~
- 0.553863 {\cal P}_{(82)}
+ 0.091478 {\cal P}_{(83)}
- 0.553863 {\cal P}_{(84)}
+ 0.953329 {\cal P}_{(85)}
\right. \right. \nonumber \\
&& \left. \left. ~~~~
- 0.157073 {\cal P}_{(86)}
+ 0.060111 {\cal P}_{(87)}
+ 0.307987 {\cal P}_{(88)}
- 0.126382 {\cal P}_{(89)}
\right. \right. \nonumber \\
&& \left. \left. ~~~~
- 0.585230 {\cal P}_{(90)}
- 0.150913 {\cal P}_{(91)}
- 0.368098 {\cal P}_{(92)}
+ 0.066271 {\cal P}_{(93)}
\right. \right. \nonumber \\
&& \left. \left. ~~~~
- 0.150913 {\cal P}_{(94)}
- 0.368098 {\cal P}_{(95)}
+ 0.066271 {\cal P}_{(96)}
+ 0.307987 {\cal P}_{(97)}
\right. \right. \nonumber \\
&& \left. \left. ~~~~
- 0.126382 {\cal P}_{(98)}
- 0.585230 {\cal P}_{(99)}
+ 0.953329 {\cal P}_{(100)}
- 0.157073 {\cal P}_{(101)}
\right. \right. \nonumber \\
&& \left. \left. ~~~~
+ 0.060111 {\cal P}_{(102)}
+ 1.184049 {\cal P}_{(103)}
+ 0.149792 {\cal P}_{(104)}
- 0.270205 {\cal P}_{(105)}
\right. \right. \nonumber \\
&& \left. \left. ~~~~
+ 0.149792 {\cal P}_{(106)}
- 0.067392 {\cal P}_{(107)}
+ 0.290831 {\cal P}_{(108)}
+ 0.454011 {\cal P}_{(109)}
\right. \right. \nonumber \\
&& \left. \left. ~~~~
- 0.447757 {\cal P}_{(110)}
+ 0.127260 {\cal P}_{(111)}
- 0.137716 {\cal P}_{(112)}
- 0.199881 {\cal P}_{(113)}
\right. \right. \nonumber \\
&& \left. \left. ~~~~
+ 0.025854 {\cal P}_{(114)}
- 0.603803 {\cal P}_{(115)}
+ 0.142944 {\cal P}_{(116)}
- 0.736292 {\cal P}_{(117)}
\right. \right. \nonumber \\
&& \left. \left. ~~~~
- 0.090950 {\cal P}_{(118)}
+ 0.275433 {\cal P}_{(119)}
- 0.090950 {\cal P}_{(120)}
- 0.736292 {\cal P}_{(121)}
\right. \right. \nonumber \\
&& \left. \left. ~~~~
+ 0.142944 {\cal P}_{(122)}
- 0.603803 {\cal P}_{(123)}
+ 0.041538 {\cal P}_{(124)}
- 0.299584 {\cal P}_{(125)}
\right. \right. \nonumber \\
&& \left. \left. ~~~~
+ 0.041538 {\cal P}_{(126)}
+ 0.127260 {\cal P}_{(127)}
- 0.447757 {\cal P}_{(128)}
+ 0.454011 {\cal P}_{(129)}
\right. \right. \nonumber \\
&& \left. \left. ~~~~
+ 0.290831 {\cal P}_{(130)}
- 0.067392 {\cal P}_{(131)}
+ 0.149792 {\cal P}_{(132)}
- 0.270205 {\cal P}_{(133)}
\right. \right. \nonumber \\
&& \left. \left. ~~~~
+ 0.149792 {\cal P}_{(134)}
+ 1.184049 {\cal P}_{(135)}
+ 0.025854 {\cal P}_{(136)}
- 0.199881 {\cal P}_{(137)}
\right. \right. \nonumber \\
&& \left. \left. ~~~~
- 0.137716 {\cal P}_{(138)}
\right] f_4^{abcd} \Nf
\right. \nonumber \\
&& \left.
+ \left[ 
5.316244 {\cal P}_{(2)}
- 5.316244 {\cal P}_{(3)}
- 1.778964 {\cal P}_{(4)}
- 3.517872 {\cal P}_{(5)}
\right. \right. \nonumber \\
&& \left. \left. ~~~~
- 1.937055 {\cal P}_{(6)}
- 1.712767 {\cal P}_{(7)}
- 0.894191 {\cal P}_{(8)}
- 1.052282 {\cal P}_{(9)}
\right. \right. \nonumber \\
&& \left. \left. ~~~~
- 2.064702 {\cal P}_{(10)}
- 1.179930 {\cal P}_{(11)}
- 3.116984 {\cal P}_{(12)}
+ 2.673156 {\cal P}_{(13)}
\right. \right. \nonumber \\
&& \left. \left. ~~~~
+ 2.989336 {\cal P}_{(14)}
+ 5.230640 {\cal P}_{(15)}
+ 3.244632 {\cal P}_{(16)}
+ 6.233969 {\cal P}_{(17)}
\right. \right. \nonumber \\
&& \left. \left. ~~~~
+ 3.244632 {\cal P}_{(18)}
+ 5.230640 {\cal P}_{(19)}
+ 2.989336 {\cal P}_{(20)}
+ 2.673156 {\cal P}_{(21)}
\right. \right. \nonumber \\
&& \left. \left. ~~~~
+ 2.557484 {\cal P}_{(22)}
+ 2.623681 {\cal P}_{(23)}
+ 1.738908 {\cal P}_{(24)}
+ 0.818576 {\cal P}_{(25)}
\right. \right. \nonumber \\
&& \left. \left. ~~~~
- 0.894191 {\cal P}_{(26)}
+ 0.158090 {\cal P}_{(27)}
- 0.066197 {\cal P}_{(28)}
+ 0.285738 {\cal P}_{(29)}
\right. \right. \nonumber \\
&& \left. \left. ~~~~
- 1.778964 {\cal P}_{(30)}
- 3.116984 {\cal P}_{(31)}
- 1.179930 {\cal P}_{(32)}
- 2.064702 {\cal P}_{(33)}
\right. \right. \nonumber \\
&& \left. \left. ~~~~
- 1.052282 {\cal P}_{(34)}
- 0.894191 {\cal P}_{(35)}
- 1.712767 {\cal P}_{(36)}
- 1.937055 {\cal P}_{(37)}
\right. \right. \nonumber \\
&& \left. \left. ~~~~
- 3.517872 {\cal P}_{(38)}
- 1.778964 {\cal P}_{(39)}
+ 2.673151 {\cal P}_{(40)}
- 2.557486 {\cal P}_{(41)}
\right. \right. \nonumber \\
&& \left. \left. ~~~~
- 0.571477 {\cal P}_{(42)}
- 2.557486 {\cal P}_{(43)}
- 5.114962 {\cal P}_{(44)}
- 2.557486 {\cal P}_{(45)}
\right. \right. \nonumber \\
&& \left. \left. ~~~~
- 0.316180 {\cal P}_{(46)}
- 2.557486 {\cal P}_{(47)}
+ 2.673151 {\cal P}_{(48)}
- 1.778964 {\cal P}_{(49)}
\right. \right. \nonumber \\
&& \left. \left. ~~~~
+ 0.285732 {\cal P}_{(50)}
- 0.066193 {\cal P}_{(51)}
+ 0.158090 {\cal P}_{(52)}
- 0.894194 {\cal P}_{(53)}
\right. \right. \nonumber \\
&& \left. \left. ~~~~
+ 0.818572 {\cal P}_{(54)}
+ 1.738904 {\cal P}_{(55)}
+ 2.623680 {\cal P}_{(56)}
+ 2.557486 {\cal P}_{(57)}
\right. \right. \nonumber \\
&& \left. \left. ~~~~
- 1.805822 {\cal P}_{(58)}
+ 3.611644 {\cal P}_{(59)}
- 1.805822 {\cal P}_{(60)}
- 3.549941 {\cal P}_{(61)}
\right. \right. \nonumber \\
&& \left. \left. ~~~~
- 3.526038 {\cal P}_{(62)}
+ 0.227060 {\cal P}_{(63)}
- 0.573359 {\cal P}_{(64)}
- 1.232465 {\cal P}_{(65)}
\right. \right. \nonumber \\
&& \left. \left. ~~~~
- 2.032884 {\cal P}_{(66)}
+ 1.720206 {\cal P}_{(67)}
+ 1.744128 {\cal P}_{(68)}
+ 4.782416 {\cal P}_{(69)}
\right. \right. \nonumber \\
&& \left. \left. ~~~~
+ 1.805822 {\cal P}_{(70)}
- 1.170762 {\cal P}_{(71)}
+ 1.805822 {\cal P}_{(72)}
- 3.549941 {\cal P}_{(73)}
\right. \right. \nonumber \\
&& \left. \left. ~~~~
- 0.573359 {\cal P}_{(74)}
+ 0.227060 {\cal P}_{(75)}
- 3.526038 {\cal P}_{(76)}
- 1.232465 {\cal P}_{(77)}
\right. \right. \nonumber \\
&& \left. \left. ~~~~
+ 1.744128 {\cal P}_{(78)}
+ 1.720208 {\cal P}_{(79)}
- 2.032883 {\cal P}_{(80)}
+ 4.782416 {\cal P}_{(81)}
\right. \right. \nonumber \\
&& \left. \left. ~~~~
+ 1.805823 {\cal P}_{(82)}
- 1.170769 {\cal P}_{(83)}
+ 1.805823 {\cal P}_{(84)}
- 3.819839 {\cal P}_{(85)}
\right. \right. \nonumber \\
&& \left. \left. ~~~~
+ 0.193345 {\cal P}_{(86)}
- 0.867162 {\cal P}_{(87)}
- 0.843246 {\cal P}_{(88)}
- 0.583153 {\cal P}_{(89)}
\right. \right. \nonumber \\
&& \left. \left. ~~~~
+ 2.109430 {\cal P}_{(90)}
+ 0.649901 {\cal P}_{(91)}
+ 1.710409 {\cal P}_{(92)}
+ 1.450315 {\cal P}_{(93)}
\right. \right. \nonumber \\
&& \left. \left. ~~~~
+ 0.649901 {\cal P}_{(94)}
+ 1.710409 {\cal P}_{(95)}
+ 1.450315 {\cal P}_{(96)}
- 0.843246 {\cal P}_{(97)}
\right. \right. \nonumber \\
&& \left. \left. ~~~~
- 0.583153 {\cal P}_{(98)}
+ 2.109430 {\cal P}_{(99)}
- 3.819839 {\cal P}_{(100)}
+ 0.193345 {\cal P}_{(101)}
\right. \right. \nonumber \\
&& \left. \left. ~~~~
- 0.867162 {\cal P}_{(102)}
- 5.621879 {\cal P}_{(103)}
- 0.081106 {\cal P}_{(104)}
+ 2.826341 {\cal P}_{(105)}
\right. \right. \nonumber \\
&& \left. \left. ~~~~
- 0.701413 {\cal P}_{(106)}
- 0.881520 {\cal P}_{(107)}
- 2.669203 {\cal P}_{(108)}
- 1.720159 {\cal P}_{(109)}
\right. \right. \nonumber \\
&& \left. \left. ~~~~
+ 1.457590 {\cal P}_{(110)}
- 1.492175 {\cal P}_{(111)}
+ 0.463316 {\cal P}_{(112)}
+ 1.481505 {\cal P}_{(113)}
\right. \right. \nonumber \\
&& \left. \left. ~~~~
+ 2.056848 {\cal P}_{(114)}
+ 1.801265 {\cal P}_{(115)}
- 1.334166 {\cal P}_{(116)}
+ 4.164290 {\cal P}_{(117)}
\right. \right. \nonumber \\
&& \left. \left. ~~~~
+ 1.187697 {\cal P}_{(118)}
- 0.926631 {\cal P}_{(119)}
+ 1.187697 {\cal P}_{(120)}
+ 4.164290 {\cal P}_{(121)}
\right. \right. \nonumber \\
&& \left. \left. ~~~~
- 1.334166 {\cal P}_{(122)}
+ 1.801265 {\cal P}_{(123)}
- 1.175328 {\cal P}_{(124)}
+ 1.402826 {\cal P}_{(125)}
\right. \right. \nonumber \\
&& \left. \left. ~~~~
- 1.175328 {\cal P}_{(126)}
- 1.492175 {\cal P}_{(127)}
+ 1.457590 {\cal P}_{(128)}
- 1.720159 {\cal P}_{(129)}
\right. \right. \nonumber \\
&& \left. \left. ~~~~
- 2.669203 {\cal P}_{(130)}
- 0.881520 {\cal P}_{(131)}
- 0.701413 {\cal P}_{(132)}
+ 2.826341 {\cal P}_{(133)}
\right. \right. \nonumber \\
&& \left. \left. ~~~~
- 0.081106 {\cal P}_{(134)}
- 5.621879 {\cal P}_{(135)}
+ 2.056848 {\cal P}_{(136)}
+ 1.481505 {\cal P}_{(137)}
\right. \right. \nonumber \\
&& \left. \left. ~~~~
+ 0.463316 {\cal P}_{(138)}
\right] f_4^{abcd} 
\right. \nonumber \\
&& \left. 
+ \left[ 
- 0.353158 {\cal P}_{(1)}
+ 0.353158 {\cal P}_{(3)}
- 0.386947 {\cal P}_{(4)}
- 0.775259 {\cal P}_{(5)}
\right. \right. \nonumber \\
&& \left. \left. ~~~~
- 0.342181 {\cal P}_{(6)}
- 0.775259 {\cal P}_{(7)}
- 0.386947 {\cal P}_{(8)}
- 0.342181 {\cal P}_{(9)}
\right. \right. \nonumber \\
&& \left. \left. ~~~~
- 0.684993 {\cal P}_{(10)}
- 0.684993 {\cal P}_{(11)}
- 1.027174 {\cal P}_{(12)}
+ 0.396705 {\cal P}_{(13)}
\right. \right. \nonumber \\
&& \left. \left. ~~~~
+ 0.374322 {\cal P}_{(14)}
- 0.086664 {\cal P}_{(15)}
+ 0.545727 {\cal P}_{(16)}
+ 0.513587 {\cal P}_{(17)}
\right. \right. \nonumber \\
&& \left. \left. ~~~~
+ 0.139265 {\cal P}_{(18)}
+ 0.861923 {\cal P}_{(19)}
- 0.032141 {\cal P}_{(20)}
- 0.009757 {\cal P}_{(21)}
\right. \right. \nonumber \\
&& \left. \left. ~~~~
- 0.388312 {\cal P}_{(22)}
- 0.465219 {\cal P}_{(23)}
- 0.871681 {\cal P}_{(24)}
+ 0.483369 {\cal P}_{(25)}
\right. \right. \nonumber \\
&& \left. \left. ~~~~
+ 0.396705 {\cal P}_{(26)}
+ 0.022383 {\cal P}_{(27)}
+ 0.076907 {\cal P}_{(28)}
- 0.149023 {\cal P}_{(29)}
\right. \right. \nonumber \\
&& \left. \left. ~~~~
- 0.009757 {\cal P}_{(30)}
+ 0.513587 {\cal P}_{(31)}
+ 0.545727 {\cal P}_{(32)}
+ 0.139265 {\cal P}_{(33)}
\right. \right. \nonumber \\
&& \left. \left. ~~~~
+ 0.374321 {\cal P}_{(34)}
+ 0.396705 {\cal P}_{(35)}
- 0.086664 {\cal P}_{(36)}
- 0.032141 {\cal P}_{(37)}
\right. \right. \nonumber \\
&& \left. \left. ~~~~
+ 0.861923 {\cal P}_{(38)}
- 0.009757 {\cal P}_{(39)}
+ 0.396705 {\cal P}_{(40)}
- 0.465219 {\cal P}_{(41)}
\right. \right. \nonumber \\
&& \left. \left. ~~~~
- 0.149023 {\cal P}_{(42)}
+ 0.483369 {\cal P}_{(43)}
- 0.388312 {\cal P}_{(44)}
+ 0.076907 {\cal P}_{(45)}
\right. \right. \nonumber \\
&& \left. \left. ~~~~
+ 0.022383 {\cal P}_{(46)}
- 0.871681 {\cal P}_{(47)}
- 0.009757 {\cal P}_{(48)}
- 0.386947 {\cal P}_{(49)}
\right. \right. \nonumber \\
&& \left. \left. ~~~~
+ 0.298046 {\cal P}_{(50)}
+ 0.388312 {\cal P}_{(51)}
- 0.044766 {\cal P}_{(52)}
- 0.386947 {\cal P}_{(53)}
\right. \right. \nonumber \\
&& \left. \left. ~~~~
+ 0.388312 {\cal P}_{(54)}
+ 0.388312 {\cal P}_{(55)}
+ 0.388312 {\cal P}_{(56)}
+ 0.776624 {\cal P}_{(57)}
\right. \right. \nonumber \\
&& \left. \left. ~~~~
+ 0.553863 {\cal P}_{(58)}
+ 0.553863 {\cal P}_{(59)}
- 1.107727 {\cal P}_{(60)}
+ 0.602682 {\cal P}_{(61)}
\right. \right. \nonumber \\
&& \left. \left. ~~~~
- 0.290535 {\cal P}_{(62)}
+ 0.385497 {\cal P}_{(63)}
- 0.042659 {\cal P}_{(64)}
+ 0.596523 {\cal P}_{(65)}
\right. \right. \nonumber \\
&& \left. \left. ~~~~
+ 0.168366 {\cal P}_{(66)}
+ 0.844399 {\cal P}_{(67)}
- 0.048819 {\cal P}_{(68)}
+ 0.596523 {\cal P}_{(69)}
\right. \right. \nonumber \\
&& \left. \left. ~~~~
+ 0.168366 {\cal P}_{(70)}
- 0.048819 {\cal P}_{(71)}
+ 0.844399 {\cal P}_{(72)}
- 1.199205 {\cal P}_{(73)}
\right. \right. \nonumber \\
&& \left. \left. ~~~~
+ 0.091478 {\cal P}_{(74)}
- 0.553863 {\cal P}_{(75)}
- 0.553863 {\cal P}_{(76)}
- 1.199205 {\cal P}_{(77)}
\right. \right. \nonumber \\
&& \left. \left. ~~~~
+ 0.091478 {\cal P}_{(78)}
- 0.553863 {\cal P}_{(79)}
- 0.553863 {\cal P}_{(80)}
+ 0.602682 {\cal P}_{(81)}
\right. \right. \nonumber \\
&& \left. \left. ~~~~
- 0.290535 {\cal P}_{(82)}
- 0.042659 {\cal P}_{(83)}
+ 0.385497 {\cal P}_{(84)}
- 0.368099 {\cal P}_{(85)}
\right. \right. \nonumber \\
&& \left. \left. ~~~~
- 0.150914 {\cal P}_{(86)}
+ 0.066271 {\cal P}_{(87)}
- 0.150914 {\cal P}_{(88)}
+ 0.066271 {\cal P}_{(89)}
\right. \right. \nonumber \\
&& \left. \left. ~~~~
- 0.368099 {\cal P}_{(90)}
- 0.157073 {\cal P}_{(91)}
+ 0.953329 {\cal P}_{(92)}
+ 0.060111 {\cal P}_{(93)}
\right. \right. \nonumber \\
&& \left. \left. ~~~~
+ 0.307987 {\cal P}_{(94)}
- 0.585230 {\cal P}_{(95)}
- 0.126382 {\cal P}_{(96)}
- 0.157073 {\cal P}_{(97)}
\right. \right. \nonumber \\
&& \left. \left. ~~~~
+ 0.060111 {\cal P}_{(98)}
+ 0.953329 {\cal P}_{(99)}
- 0.585230 {\cal P}_{(100)}
+ 0.307987 {\cal P}_{(101)}
\right. \right. \nonumber \\
&& \left. \left. ~~~~
- 0.126382 {\cal P}_{(102)}
- 0.447757 {\cal P}_{(103)}
+ 0.454011 {\cal P}_{(104)}
+ 0.127261 {\cal P}_{(105)}
\right. \right. \nonumber \\
&& \left. \left. ~~~~
+ 0.149792 {\cal P}_{(106)}
+ 0.025855 {\cal P}_{(107)}
- 0.199881 {\cal P}_{(108)}
+ 0.149792 {\cal P}_{(109)}
\right. \right. \nonumber \\
&& \left. \left. ~~~~
+ 1.184049 {\cal P}_{(110)}
- 0.270205 {\cal P}_{(111)}
- 0.137716 {\cal P}_{(112)}
+ 0.290832 {\cal P}_{(113)}
\right. \right. \nonumber \\
&& \left. \left. ~~~~
- 0.067392 {\cal P}_{(114)}
+ 0.149792 {\cal P}_{(115)}
- 0.270205 {\cal P}_{(116)}
+ 1.184049 {\cal P}_{(117)}
\right. \right. \nonumber \\
&& \left. \left. ~~~~
+ 0.290832 {\cal P}_{(118)}
- 0.137716 {\cal P}_{(119)}
- 0.199881 {\cal P}_{(120)}
- 0.447757 {\cal P}_{(121)}
\right. \right. \nonumber \\
&& \left. \left. ~~~~
+ 0.127261 {\cal P}_{(122)}
+ 0.454011 {\cal P}_{(123)}
+ 0.025855 {\cal P}_{(124)}
+ 0.149792 {\cal P}_{(125)}
\right. \right. \nonumber \\
&& \left. \left. ~~~~
- 0.067392 {\cal P}_{(126)}
+ 0.142944 {\cal P}_{(127)}
- 0.736292 {\cal P}_{(128)}
- 0.603804 {\cal P}_{(129)}
\right. \right. \nonumber \\
&& \left. \left. ~~~~
- 0.090951 {\cal P}_{(130)}
+ 0.041538 {\cal P}_{(131)}
- 0.299585 {\cal P}_{(132)}
+ 0.142944 {\cal P}_{(133)}
\right. \right. \nonumber \\
&& \left. \left. ~~~~
- 0.603804 {\cal P}_{(134)}
- 0.736292 {\cal P}_{(135)}
+ 0.041538 {\cal P}_{(136)}
- 0.090951 {\cal P}_{(137)}
\right. \right. \nonumber \\
&& \left. \left. ~~~~
+ 0.275433 {\cal P}_{(138)}
\right] f_4^{acbd} \Nf
\right. \nonumber \\
&& \left. 
+ \left[ 
5.316244 {\cal P}_{(1)}
- 5.316244 {\cal P}_{(3)}
+ 2.673156 {\cal P}_{(4)}
+ 5.230640 {\cal P}_{(5)}
\right. \right. \nonumber \\
&& \left. \left. ~~~~
+ 2.989336 {\cal P}_{(6)}
+ 5.230640 {\cal P}_{(7)}
+ 2.673156 {\cal P}_{(8)}
+ 2.989336 {\cal P}_{(9)}
\right. \right. \nonumber \\
&& \left. \left. ~~~~
+ 3.244632 {\cal P}_{(10)}
+ 3.244632 {\cal P}_{(11)}
+ 6.233969 {\cal P}_{(12)}
- 1.778964 {\cal P}_{(13)}
\right. \right. \nonumber \\
&& \left. \left. ~~~~
- 1.937055 {\cal P}_{(14)}
- 3.517872 {\cal P}_{(15)}
- 2.064702 {\cal P}_{(16)}
- 3.116984 {\cal P}_{(17)}
\right. \right. \nonumber \\
&& \left. \left. ~~~~
- 1.179930 {\cal P}_{(18)}
- 1.712767 {\cal P}_{(19)}
- 1.052282 {\cal P}_{(20)}
- 0.894191 {\cal P}_{(21)}
\right. \right. \nonumber \\
&& \left. \left. ~~~~
+ 2.557484 {\cal P}_{(22)}
- 0.066197 {\cal P}_{(23)}
+ 0.818576 {\cal P}_{(24)}
+ 1.738908 {\cal P}_{(25)}
\right. \right. \nonumber \\
&& \left. \left. ~~~~
- 1.778964 {\cal P}_{(26)}
+ 0.158090 {\cal P}_{(27)}
+ 2.623681 {\cal P}_{(28)}
+ 0.285738 {\cal P}_{(29)}
\right. \right. \nonumber \\
&& \left. \left. ~~~~
- 0.894191 {\cal P}_{(30)}
- 3.116984 {\cal P}_{(31)}
- 2.064702 {\cal P}_{(32)}
- 1.179930 {\cal P}_{(33)}
\right. \right. \nonumber \\
&& \left. \left. ~~~~
- 1.937055 {\cal P}_{(34)}
- 1.778964 {\cal P}_{(35)}
- 3.517872 {\cal P}_{(36)}
- 1.052282 {\cal P}_{(37)}
\right. \right. \nonumber \\
&& \left. \left. ~~~~
- 1.712767 {\cal P}_{(38)}
- 0.894191 {\cal P}_{(39)}
- 1.778964 {\cal P}_{(40)}
- 0.066197 {\cal P}_{(41)}
\right. \right. \nonumber \\
&& \left. \left. ~~~~
+ 0.285738 {\cal P}_{(42)}
+ 1.738908 {\cal P}_{(43)}
+ 2.557484 {\cal P}_{(44)}
+ 2.623681 {\cal P}_{(45)}
\right. \right. \nonumber \\
&& \left. \left. ~~~~
+ 0.158090 {\cal P}_{(46)}
+ 0.818576 {\cal P}_{(47)}
- 0.894191 {\cal P}_{(48)}
+ 2.673156 {\cal P}_{(49)}
\right. \right. \nonumber \\
&& \left. \left. ~~~~
- 0.571476 {\cal P}_{(50)}
- 2.557484 {\cal P}_{(51)}
- 0.316181 {\cal P}_{(52)}
+ 2.673156 {\cal P}_{(53)}
\right. \right. \nonumber \\
&& \left. \left. ~~~~
- 2.557484 {\cal P}_{(54)}
- 2.557484 {\cal P}_{(55)}
- 2.557484 {\cal P}_{(56)}
- 5.114968 {\cal P}_{(57)}
\right. \right. \nonumber \\
&& \left. \left. ~~~~
- 1.805823 {\cal P}_{(58)}
- 1.805823 {\cal P}_{(59)}
+ 3.611646 {\cal P}_{(60)}
- 1.232469 {\cal P}_{(61)}
\right. \right. \nonumber \\
&& \left. \left. ~~~~
+ 1.720208 {\cal P}_{(62)}
- 2.032883 {\cal P}_{(63)}
+ 1.744124 {\cal P}_{(64)}
- 3.549947 {\cal P}_{(65)}
\right. \right. \nonumber \\
&& \left. \left. ~~~~
+ 0.227060 {\cal P}_{(66)}
- 3.526031 {\cal P}_{(67)}
- 0.573354 {\cal P}_{(68)}
- 3.549947 {\cal P}_{(69)}
\right. \right. \nonumber \\
&& \left. \left. ~~~~
+ 0.227060 {\cal P}_{(70)}
- 0.573354 {\cal P}_{(71)}
- 3.526030 {\cal P}_{(72)}
+ 4.782416 {\cal P}_{(73)}
\right. \right. \nonumber \\
&& \left. \left. ~~~~
- 1.170770 {\cal P}_{(74)}
+ 1.805823 {\cal P}_{(75)}
+ 1.805823 {\cal P}_{(76)}
+ 4.782416 {\cal P}_{(77)}
\right. \right. \nonumber \\
&& \left. \left. ~~~~
- 1.170770 {\cal P}_{(78)}
+ 1.805823 {\cal P}_{(79)}
+ 1.805823 {\cal P}_{(80)}
- 1.232470 {\cal P}_{(81)}
\right. \right. \nonumber \\
&& \left. \left. ~~~~
+ 1.720208 {\cal P}_{(82)}
+ 1.744124 {\cal P}_{(83)}
- 2.032883 {\cal P}_{(84)}
+ 1.710409 {\cal P}_{(85)}
\right. \right. \nonumber \\
&& \left. \left. ~~~~
+ 0.649901 {\cal P}_{(86)}
+ 1.450315 {\cal P}_{(87)}
+ 0.649901 {\cal P}_{(88)}
+ 1.450315 {\cal P}_{(89)}
\right. \right. \nonumber \\
&& \left. \left. ~~~~
+ 1.710409 {\cal P}_{(90)}
+ 0.193345 {\cal P}_{(91)}
- 3.819839 {\cal P}_{(92)}
- 0.867162 {\cal P}_{(93)}
\right. \right. \nonumber \\
&& \left. \left. ~~~~
- 0.843246 {\cal P}_{(94)}
+ 2.109430 {\cal P}_{(95)}
- 0.583153 {\cal P}_{(96)}
+ 0.193345 {\cal P}_{(97)}
\right. \right. \nonumber \\
&& \left. \left. ~~~~
- 0.867162 {\cal P}_{(98)}
- 3.819839 {\cal P}_{(99)}
+ 2.109430 {\cal P}_{(100)}
- 0.843246 {\cal P}_{(101)}
\right. \right. \nonumber \\
&& \left. \left. ~~~~
- 0.583153 {\cal P}_{(102)}
+ 1.457590 {\cal P}_{(103)}
- 1.720159 {\cal P}_{(104)}
- 1.492175 {\cal P}_{(105)}
\right. \right. \nonumber \\
&& \left. \left. ~~~~
- 0.701412 {\cal P}_{(106)}
+ 2.056848 {\cal P}_{(107)}
+ 1.481505 {\cal P}_{(108)}
- 0.081106 {\cal P}_{(109)}
\right. \right. \nonumber \\
&& \left. \left. ~~~~
- 5.621879 {\cal P}_{(110)}
+ 2.826341 {\cal P}_{(111)}
+ 0.463316 {\cal P}_{(112)}
- 2.669203 {\cal P}_{(113)}
\right. \right. \nonumber \\
&& \left. \left. ~~~~
- 0.881520 {\cal P}_{(114)}
- 0.081106 {\cal P}_{(115)}
+ 2.826341 {\cal P}_{(116)}
- 5.621879 {\cal P}_{(117)}
\right. \right. \nonumber \\
&& \left. \left. ~~~~
- 2.669203 {\cal P}_{(118)}
+ 0.463316 {\cal P}_{(119)}
+ 1.481505 {\cal P}_{(120)}
+ 1.457590 {\cal P}_{(121)}
\right. \right. \nonumber \\
&& \left. \left. ~~~~
- 1.492175 {\cal P}_{(122)}
- 1.720159 {\cal P}_{(123)}
+ 2.056848 {\cal P}_{(124)}
- 0.701413 {\cal P}_{(125)}
\right. \right. \nonumber \\
&& \left. \left. ~~~~
- 0.881520 {\cal P}_{(126)}
- 1.334166 {\cal P}_{(127)}
+ 4.164290 {\cal P}_{(128)}
+ 1.801265 {\cal P}_{(129)}
\right. \right. \nonumber \\
&& \left. \left. ~~~~
+ 1.187697 {\cal P}_{(130)}
- 1.175328 {\cal P}_{(131)}
+ 1.402826 {\cal P}_{(132)}
- 1.334166 {\cal P}_{(133)}
\right. \right. \nonumber \\
&& \left. \left. ~~~~
+ 1.801265 {\cal P}_{(134)}
+ 4.164290 {\cal P}_{(135)}
- 1.175328 {\cal P}_{(136)}
+ 1.187697 {\cal P}_{(137)}
\right. \right. \nonumber \\
&& \left. \left. ~~~~
- 0.926631 {\cal P}_{(138)}
\right] f_4^{acbd} 
+ \left[ 
- 0.353158 {\cal P}_{(1)}
+ 0.353158{\cal P}_{(2)}
\right] f_4^{adbc} \Nf  
\right. \nonumber \\
&& \left. 
+ \left[
5.316244 {\cal P}_{(1)}
- 5.316244 {\cal P}_{(2)}
\right] f_4^{adbc} 
\right] a ~+~ O(a^2) 
\end{eqnarray} 
where we have dropped the common argument of ${\cal P}_{(k)}(p,q,r)$ and have
only used the Jacobi identity in the channels with a non-zero tree term. Indeed
for these three channels, $1$, $2$ and $3$, the symmetry associated with colour
tensors $d_F^{abcd}$ and $d_A^{abcd}$ is evident. 

\sect{$\MOMgggg$ scheme.}

Having completely determined the $4$-point function at one loop in the 
$\MSbar$ scheme at the symmetric point we can now consider the renormalization
in other schemes. In \cite{16} the symmetric point renormalization of the
$3$-point vertices led naturally into the definition of the momentum
subtraction schemes. These are mass dependent schemes and are constructed in
such a way that after renormalization at the subtraction point the respective
vertices have no $O(a)$ corrections. This is in addition to the wave function
renormalization constants being defined in the same way via the $2$-point
functions. As there are three $3$-point vertices in QCD this leads to three
separate MOM schemes which are denoted by $\MOMggg$, $\MOMh$ and $\MOMq$ based 
on the respective triple gluon, ghost-gluon and quark-gluon vertices,
\cite{16}. In light of this and the fact that we have considered the quartic 
gluon vertex at the fully symmetric point we can define an analogous momentum 
subtraction scheme which will be denoted by $\MOMgggg$. More specifically the 
scheme is defined as follows. We will retain the wave function MOM 
renormalization scheme approach. By this we mean that the $2$-point functions 
are rendered finite by ensuring that at the subtraction point there are no 
$O(a)$ corrections. This does not mean that the wave function renormalization 
constants are the same as the three MOM schemes of \cite{16}. This is because
as one proceeds beyond one loop the renormalization constant of the coupling 
constant is required and this is different in different MOM schemes. Therefore,
that observation will equally apply to $\MOMgggg$. However, for the $\MOMgggg$ 
coupling constant renormalization its definition requires some care. This is 
because unlike the $3$-point vertices there is more than one colour tensor 
structure for the $4$-point function. Therefore, we require the quartic vertex 
function to be written in terms of the colour tensors of the original quartic 
gluon Feynman rule for $SU(\Nc)$ in contrast to \cite{18}. This was one of the 
reasons for already presenting our results in this format in the previous 
section. Thus the $\MOMgggg$ scheme coupling constant renormalization is 
defined so that after renormalization there are no $O(a)$ corrections to that 
part of the vertex function corresponding to the original quartic vertex 
Feynman rule. With this definition we find the coupling constant 
renormalization constant is
\begin{eqnarray}
Z_g^{\MOMggggs} &=& 1 + \left[ \frac{1}{3} \Nf - \frac{11}{6} \Nc \right] 
\frac{a}{\epsilon} \nonumber \\
&& + \left[ 
- \frac{5329}{1440} 
- \frac{17}{32} \alpha 
+ \frac{53}{160} \alpha^2
- \frac{7}{160} \alpha^3 
- \frac{431}{150} \ln \left( \frac{4}{3} \right) 
+ \frac{27}{160} \ln \left( \frac{4}{3} \right) \alpha
\right. \nonumber \\
&& \left. ~~~
+ \frac{253}{400} \ln \left( \frac{4}{3} \right) \alpha^2 
- \frac{9}{800} \ln \left( \frac{4}{3} \right) \alpha^3
+ \frac{7219}{76800} \Phi_1 \left( \frac{9}{16}, \frac{9}{16} \right)
\right. \nonumber \\
&& \left. ~~~
+ \frac{897}{5120} \Phi_1 \left( \frac{9}{16}, \frac{9}{16} \right) \alpha
- \frac{2709}{25600} \Phi_1 \left( \frac{9}{16}, \frac{9}{16} \right) \alpha^2
+ \frac{351}{25600} \Phi_1 \left( \frac{9}{16}, \frac{9}{16} \right) \alpha^3 
\right. \nonumber \\
&& \left. ~~~
+ \frac{85}{384} \Phi_1 \left( \frac{3}{4}, \frac{3}{4} \right)
+ \frac{5}{64} \Phi_1 \left( \frac{3}{4}, \frac{3}{4} \right) \alpha 
- \frac{9}{128} \Phi_1 \left( \frac{3}{4}, \frac{3}{4} \right) \alpha^2 
\right] \Nc \nonumber \\
&&
+ \left[ 
\frac{7}{9} 
+ \frac{7}{12} \ln \left( \frac{4}{3} \right) 
- \frac{1}{48} \Phi_1 \left( \frac{9}{16}, \frac{9}{16} \right)
+ \frac{7}{48} \Phi_1 \left( \frac{3}{4}, \frac{3}{4} \right) 
\right] \Nf a ~+~ O(a^2) ~.
\label{zgmom}
\end{eqnarray} 
Equipped with this renormalization constant we can construct the relation 
between the $\Lambda$-parameters in $\MOMgggg$ and $\MSbar$. From standard 
formalism this ratio is defined as 
\begin{equation}
\frac{\Lambda^{\MOMggggs}}{\Lambda^{\MSbars}} ~=~ \exp \left[
\frac{\lambda^{\MOMggggs}(\alpha,\Nf)}{b_0} \right]
\end{equation}
where we take
\begin{equation}
b_0 ~=~ \frac{22}{3} \Nc ~-~ \frac{4}{3} \Nf
\end{equation}
and $\lambda^{\MOMggggs}(\alpha,\Nf)$ is related to the finite part of the one
loop coupling constant renormalization constant. From $Z_g^{\MOMggggs}$ we have 
\begin{eqnarray}
\lambda^{\MOMggggs}(\alpha,\Nf) &=&
\frac{1}{115200}
\left[ 2592 \ln \left( \frac{4}{3} \right) \alpha^3 \Nc 
- 145728 \ln \left( \frac{4}{3} \right) \alpha^2 \Nc
- 38880 \ln \left( \frac{4}{3} \right) \alpha \Nc 
\right. \nonumber \\
&& \left. ~~~~~~~~~~~
+ 662016 \ln \left( \frac{4}{3} \right) \Nc
- 134400 \ln \left( \frac{4}{3} \right) \Nf 
\right. \nonumber \\
&& \left. ~~~~~~~~~~~
- 3159 \Phi_1 \left( \frac{9}{16}, \frac{9}{16} \right) \alpha^3 \Nc
+ 24381 \Phi_1 \left( \frac{9}{16}, \frac{9}{16} \right) \alpha^2 \Nc
\right. \nonumber \\
&& \left. ~~~~~~~~~~~
- 40365 \Phi_1 \left( \frac{9}{16}, \frac{9}{16} \right) \alpha \Nc 
- 21657 \Phi_1 \left( \frac{9}{16}, \frac{9}{16} \right) \Nc
\right. \nonumber \\
&& \left. ~~~~~~~~~~~
+ 4800 \Phi_1 \left( \frac{9}{16}, \frac{9}{16} \right) \Nf 
+ 16200 \Phi_1 \left( \frac{3}{4}, \frac{3}{4} \right) \alpha^2 \Nc
\right. \nonumber \\
&& \left. ~~~~~~~~~~~
- 18000 \Phi_1 \left( \frac{3}{4}, \frac{3}{4} \right) \alpha \Nc 
- 51000 \Phi_1 \left( \frac{3}{4}, \frac{3}{4} \right) \Nc
\right. \nonumber \\
&& \left. ~~~~~~~~~~~
- 33600 \Phi_1 \left( \frac{3}{4}, \frac{3}{4} \right) \Nf 
+ 10080 \alpha^3 \Nc - 76320 \alpha^2 \Nc
\right. \nonumber \\
&& \left. ~~~~~~~~~~~
+ 122400 \alpha \Nc + 852640 \Nc - 179200 \Nf \right] ~.
\end{eqnarray}
To appreciate the behaviour of $\lambda^{\MOMggggs}(\alpha,\Nf)$ we have 
computed it for various values of $\alpha$ and $\Nf$ and presented the results
in Table $1$. There we reproduce the corresponding values for the three MOM
schemes defined in \cite{16} in relation to $\MSbar$ rather than the original
minimal subtraction (MS) scheme values which were actually given there. For 
instance, the Landau gauge values are lower than those for $\MOMggg$. 

{\begin{table}[ht]
\begin{center}
\begin{tabular}{|c||c||c|c|c|c|}
\hline
$\alpha$ & $\Nf$ & $\MOMgggg$ & $\MOMggg$ & $\MOMh$ & $\MOMq$ \\
\hline
0 & 0 & 2.6551 & 3.3341 & 2.3236 & 2.1379 \\
0 & 1 & 2.4965 & 3.0543 & 2.3250 & 2.1277 \\
0 & 2 & 2.3274 & 2.7644 & 2.3267 & 2.1163 \\
0 & 3 & 2.1474 & 2.4654 & 2.3286 & 2.1032 \\
0 & 4 & 1.9560 & 2.1587 & 2.3308 & 2.0881 \\
0 & 5 & 1.7529 & 1.8471 & 2.3336 & 2.0706 \\
1 & 0 & 2.4543 & 2.8957 & 2.6166 & 1.9075 \\
1 & 3 & 1.9506 & 2.0751 & 2.6924 & 1.8296 \\
1 & 4 & 1.7631 & 1.7921 & 2.7265 & 1.7964 \\
1 & 5 & 1.5658 & 1.5088 & 2.7670 & 1.7581 \\
3 & 3 & 1.7693 & 1.8392 & 4.1918 & 1.3110 \\
3 & 4 & 1.5868 & 1.5732 & 4.3978 & 1.2533 \\
-2 & 4 & 2.6576 & 2.5437 & 2.0081 & 2.6597 \\
\hline
\end{tabular}
\end{center}
\begin{center}
{Table $1$. Values of $\Lambda^{\MOMggggs}/\Lambda^{\MSbars}$ for $SU(3)$ in 
comparison with other MOM schemes of \cite{16}.}
\end{center}
\end{table}}

While we have introduced the new scheme $\MOMgggg$ we have not presented the
associated one loop renormalization group functions for this scheme. This is
because at this order these functions are scheme independent. Beyond one loop
the coefficients of each term in the coupling constant expansion depend on the
scheme. This applies to the $\beta$-function too. Though in mass independent
renormalization schemes such as $\MSbar$ the two loop term of the 
$\beta$-function is also scheme independent in theories with only one coupling
constant, \cite{47}. However, while we have performed a one loop computation it
is possible to construct the two loop renormalization group functions in the
$\MOMgggg$ scheme using properties of the renormalization group. See, for
example, \cite{48} for background to this. To achieve this we have to compute
the various conversion functions for each renormalization group function which
in essence are the ratios of the respective renormalization constants in each
scheme. In particular
\begin{eqnarray}
C^{\MOMggggs}_g(a,\alpha) &=& \frac{Z_g^{\MOMggggs}}{Z_g^{\MSbars}} ~~~~,~~~~
C^{\MOMggggs}_\phi(a,\alpha) ~=~ \frac{Z_\phi^{\MOMggggs}}{Z_\phi^{\MSbars}}
\nonumber \\
C^{\MOMggggs}_\alpha(a,\alpha) &=& \frac{Z_\alpha^{\MOMggggs}Z_A^{\MSbars}}
{Z_\alpha^{\MSbars}Z_A^{\MOMggggs}}
\end{eqnarray}
where $\phi$~$\in$~$\{A,c,\psi\}$ denote the gluon, ghost and quark fields
respectively. While in our conventions $Z_\alpha$ will be unity in a linear 
covariant gauge we have included it here so as to be formally correct. There
are (nonlinear covariant) gauges where the corresponding gauge parameter 
renormalization constant is not unity. It is important to realise that in these
formal definitions the perturbative expansion is in powers of the coupling 
constant defined in one scheme. (Our convention is that that scheme is
$\MSbar$.) Otherwise the conversion functions would have poles in $\epsilon$. 
Therefore, we have to relate the $\MOMgggg$ definition of the coupling constant
to the $\MSbar$ one order by order. This is achieved by
\begin{equation}
a_{\mbox{$\MOMggggs$}} ~=~ 
\frac{a_{\mbox{$\MSbars$}}}{\left( C^{\MOMggggs}_g(a,\alpha) \right)^2} ~. 
\end{equation}
For the $\MOMgggg$ scheme the explicit form of (\ref{zgmom}) gives  
\begin{eqnarray}
a_{\mbox{$\MOMggggs$}} &=&
a + \left[ 2592 \ln \left( \frac{4}{3} \right) \alpha^3 \Nc
- 145728 \ln \left( \frac{4}{3} \right) \alpha^2 \Nc
- 38880 \ln \left( \frac{4}{3} \right) \alpha \Nc
\right. \nonumber \\
&& \left. ~~~~~~
+ 662016 \ln \left( \frac{4}{3} \right) \Nc
- 134400 \ln \left( \frac{4}{3} \right) \Nf
- 3159 \Phi_1 \left( \frac{9}{16}, \frac{9}{16} \right) \alpha^3 \Nc
\right. \nonumber \\
&& \left. ~~~~~~
+ 24381 \Phi_1 \left( \frac{9}{16}, \frac{9}{16} \right) \alpha^2 \Nc
- 40365 \Phi_1 \left( \frac{9}{16}, \frac{9}{16} \right) \alpha \Nc
\right. \nonumber \\
&& \left. ~~~~~~
- 21657 \Phi_1 \left( \frac{9}{16}, \frac{9}{16} \right) \Nc
+ 4800 \Phi_1 \left( \frac{9}{16}, \frac{9}{16} \right) \Nf
\right. \nonumber \\
&& \left. ~~~~~~
+ 16200 \Phi_1 \left( \frac{3}{4}, \frac{3}{4} \right) \alpha^2 \Nc
- 18000 \Phi_1 \left( \frac{3}{4}, \frac{3}{4} \right) \alpha \Nc
- 51000 \Phi_1 \left( \frac{3}{4}, \frac{3}{4} \right) \Nc
\right. \nonumber \\
&& \left. ~~~~~~
- 33600 \Phi_1 \left( \frac{3}{4}, \frac{3}{4} \right) \Nf
+ 10080 \alpha^3 \Nc
- 76320 \alpha^2 \Nc
+ 122400 \alpha \Nc
\right. \nonumber \\
&& \left. ~~~~~~
+ 852640 \Nc
- 179200 \Nf \right] \frac{a^2}{115200} ~+~ O(a^3) 
\end{eqnarray} 
where the one loop correction coefficient is in effect 
$\lambda^{\MOMggggs}(\alpha,\Nf)$. In addition from the one loop wave function
renormalization constants, \cite{16}, we have  
\begin{eqnarray}
C_A^{\MOMggggs} (a,\alpha) &=&
1 + \left[ 9 \alpha^2 \Nc + 18 \alpha \Nc + 97 \Nc - 40 \Nf \right] 
\frac{a}{36} ~+~ O(a^2) \nonumber \\ 
C_c^{\MOMggggs} (a,\alpha) &=& 1 + \Nc a ~+~ O(a^2) \nonumber \\
C_\psi^{\MOMggggs} (a,\alpha) &=&
1 + \alpha \left[ - \Nc^2 + 1 \right] \frac{a}{2 \Nc} ~+~ O(a^2) ~.
\end{eqnarray}
Once these are known explicitly the two loop $\MOMgggg$ corrections to the
respective renormalization group functions can be computed from the formal
relations 
\begin{eqnarray}
\beta^{\mbox{$\MOMggggs$}} ( a_{\mbox{$\MOMggggs$}}, 
\alpha_{\mbox{$\MOMggggs$}} ) &=&
\left[ \beta^{\mbox{$\MSbars$}}( a_{\mbox{$\MSbars$}} )
\frac{\partial a_{\mbox{$\MOMggggs$}}}{\partial a_{\mbox{$\MSbars$}}}
\right. \nonumber \\
&& \left. ~
+ \alpha_{\mbox{$\MSbars$}} \gamma^{\mbox{$\MSbars$}}_\alpha
( a_{\mbox{$\MSbars$}}, \alpha_{\mbox{\footnotesize{$\MSbars$}}} )
\frac{\partial a_{\mbox{$\MOMggggs$}}}{\partial \alpha_{\mbox{$\MSbars$}}}
\right]_{ \MSbars \rightarrow \MOMggggs }
\label{betacon}
\end{eqnarray}
and
\begin{eqnarray}
\gamma_\phi^{\MOMggggs} ( a_{\MOMggggs}, \alpha_{\MOMggggs} )
&=& \!\!\! \left[ 
\beta^{\MSbars}\left(a_{\MSbars}\right)
\frac{\partial ~}{\partial a_{\MSbars}} 
\ln C_\phi^{\MOMggggs} (a_{\MSbars},\alpha_{\MSbars})
\right. \nonumber \\
&& \left.
+~ \alpha_{\MSbars} \gamma^{\MSbars}_\alpha
(a_{\MSbars},\alpha_{\MSbars})
\frac{\partial ~}{\partial \alpha_{\MSbars}} 
\ln C_\phi^{\MOMggggs} (a_{\MSbars},\alpha_{\MSbars})
\right. \nonumber \\
&& \left. 
+~ \gamma_\phi^{\MSbars} (a_{\MSbars}) 
\right]_{ \MSbars \rightarrow \MOMggggs } 
\label{gammacon}
\end{eqnarray}
which are constructed from the renormalization group equation itself. In 
(\ref{betacon}) and (\ref{gammacon}) we have labelled the variables with the
scheme they are defined in. In addition to having a coupling constant defined
in a scheme the gauge parameter is also defined with respect to a scheme. 
Though from the construction the Landau gauge is preserved between schemes. 
Also the renormalization group functions are labelled with a scheme. However, 
the mapping, ${}_{ \MSbars \rightarrow \MOMggggs }$, in (\ref{betacon}) and
(\ref{gammacon}) indicates that after the left hand side has been determined in
$\MSbar$ variables then they are mapped to their $\MOMgggg$ counterparts as 
indicated by the arguments on the right hand side. From these expressions one 
can see that to construct the two loop corrections only the one loop conversion
functions are required since each term involving such a function is multiplied 
by a function which is $O(a)$. Further, as the $\MSbar$ two loop 
renormalization group functions are available, \cite{49,50,51,52,53,54}, then 
we have all the required information to extract the {\em two} loop
renormalization group functions. 

Therefore, using this formalism and the explicit values for the conversion
functions we have  
\begin{eqnarray}
\gamma_A^{\mbox{$\MOMggggs$}}(a,\alpha) &=&
\left[ 3 \alpha \Nc - 13 \Nc + 4 \Nf \right] \frac{a}{6} \nonumber \\ 
&& + \left[ -~ 7776 \ln \left( \frac{4}{3} \right) \alpha^4 \Nc^3
+ 470880 \ln \left( \frac{4}{3} \right) \alpha^3 \Nc^3
\right. \nonumber \\
&& \left. ~~~
- 10368 \ln \left( \frac{4}{3} \right) \alpha^3 \Nc^2 \Nf
- 1777824 \ln \left( \frac{4}{3} \right) \alpha^2 \Nc^3
\right. \nonumber \\
&& \left. ~~~
+ 582912 \ln \left( \frac{4}{3} \right) \alpha^2 \Nc^2 \Nf
- 2491488 \ln \left( \frac{4}{3} \right) \alpha \Nc^3
\right. \nonumber \\
&& \left. ~~~
+ 558720 \ln \left( \frac{4}{3} \right) \alpha \Nc^2 \Nf
+ 8606208 \ln \left( \frac{4}{3} \right) \Nc^3
\right. \nonumber \\
&& \left. ~~~
- 4395264 \ln \left( \frac{4}{3} \right) \Nc^2 \Nf
+ 537600 \ln \left( \frac{4}{3} \right) \Nc \Nf^2
\right. \nonumber \\
&& \left. ~~~
+ 9477 \Phi_1\left( \frac{9}{16}, \frac{9}{16} \right) \alpha^4 \Nc^3
- 114210 \Phi_1\left( \frac{9}{16}, \frac{9}{16} \right) \alpha^3 \Nc^3
\right. \nonumber \\
&& \left. ~~~
+ 12636 \Phi_1\left( \frac{9}{16}, \frac{9}{16} \right) \alpha^3 \Nc^2 \Nf
+ 438048 \Phi_1\left( \frac{9}{16}, \frac{9}{16} \right) \alpha^2 \Nc^3
\right. \nonumber \\
&& \left. ~~~
- 97524 \Phi_1\left( \frac{9}{16}, \frac{9}{16} \right) \alpha^2 \Nc^2 \Nf
- 459774 \Phi_1\left( \frac{9}{16}, \frac{9}{16} \right) \alpha \Nc^3
\right. \nonumber \\
&& \left. ~~~
+ 147060 \Phi_1\left( \frac{9}{16}, \frac{9}{16} \right) \alpha \Nc^2 \Nf
- 281541 \Phi_1\left( \frac{9}{16}, \frac{9}{16} \right) \Nc^3
\right. \nonumber \\
&& \left. ~~~
+ 149028 \Phi_1\left( \frac{9}{16}, \frac{9}{16} \right) \Nc^2 \Nf
- 19200 \Phi_1\left( \frac{9}{16}, \frac{9}{16} \right) \Nc \Nf^2
\right. \nonumber \\
&& \left. ~~~
- 48600 \Phi_1\left( \frac{3}{4}, \frac{3}{4} \right) \alpha^3 \Nc^3
+ 264600 \Phi_1\left( \frac{3}{4}, \frac{3}{4} \right) \alpha^2 \Nc^3
\right. \nonumber \\
&& \left. ~~~
- 64800 \Phi_1\left( \frac{3}{4}, \frac{3}{4} \right) \alpha^2 \Nc^2 \Nf
- 81000 \Phi_1\left( \frac{3}{4}, \frac{3}{4} \right) \alpha \Nc^3
\right. \nonumber \\
&& \left. ~~~
+ 172800 \Phi_1\left( \frac{3}{4}, \frac{3}{4} \right) \alpha \Nc^2 \Nf
- 663000 \Phi_1\left( \frac{3}{4}, \frac{3}{4} \right) \Nc^3
\right. \nonumber \\
&& \left. ~~~
- 232800 \Phi_1\left( \frac{3}{4}, \frac{3}{4} \right) \Nc^2 \Nf
+ 134400 \Phi_1\left( \frac{3}{4}, \frac{3}{4} \right) \Nc \Nf^2
\right. \nonumber \\
&& \left. ~~~
- 30240 \alpha^4 \Nc^3
+ 273600 \alpha^3 \Nc^3
- 40320 \alpha^3 \Nc^2 \Nf
- 1071360 \alpha^2 \Nc^3
\right. \nonumber \\
&& \left. ~~~
+ 190080 \alpha^2 \Nc^2 \Nf
+ 396480 \alpha \Nc^3
- 336000 \alpha \Nc^2 \Nf
- 842080 \Nc^3
\right. \nonumber \\
&& \left. ~~~
+ 736640 \Nc^2 \Nf
+ 204800 \Nc \Nf^2
- 691200 \Nf \right] \frac{a^2}{691200 \Nc} ~+~ O(a^3) \\
\gamma_c^{\mbox{$\MOMggggs$}}(a,\alpha) &=&
[\alpha - 3] \frac{\Nc a}{4} \nonumber \\ 
&& + \left[ -~ 2592 \ln \left( \frac{4}{3} \right) \alpha^4 \Nc
+ 153504 \ln \left( \frac{4}{3} \right) \alpha^3 \Nc
- 398304 \ln \left( \frac{4}{3} \right) \alpha^2 \Nc
\right. \nonumber \\
&& \left. ~~~
- 778656 \ln \left( \frac{4}{3} \right) \alpha \Nc
+ 134400 \ln \left( \frac{4}{3} \right) \alpha \Nf
+ 1986048 \ln \left( \frac{4}{3} \right) \Nc
\right. \nonumber \\
&& \left. ~~~
- 403200 \ln \left( \frac{4}{3} \right) \Nf
+ 3159 \Phi_1\left( \frac{9}{16}, \frac{9}{16} \right) \alpha^4 \Nc
\right. \nonumber \\
&& \left. ~~~
- 33858 \Phi_1\left( \frac{9}{16}, \frac{9}{16} \right) \alpha^3 \Nc
+ 113508 \Phi_1\left( \frac{9}{16}, \frac{9}{16} \right) \alpha^2 \Nc
\right. \nonumber \\
&& \left. ~~~
- 99438 \Phi_1\left( \frac{9}{16}, \frac{9}{16} \right) \alpha \Nc
- 4800 \Phi_1\left( \frac{9}{16}, \frac{9}{16} \right) \alpha \Nf
\right. \nonumber \\
&& \left. ~~~
- 64971 \Phi_1\left( \frac{9}{16}, \frac{9}{16} \right) \Nc
+ 14400 \Phi_1\left( \frac{9}{16}, \frac{9}{16} \right) \Nf
\right. \nonumber \\
&& \left. ~~~
- 16200 \Phi_1\left( \frac{3}{4}, \frac{3}{4} \right) \alpha^3 \Nc
+ 66600 \Phi_1\left( \frac{3}{4}, \frac{3}{4} \right) \alpha^2 \Nc
\right. \nonumber \\
&& \left. ~~~
- 3000 \Phi_1\left( \frac{3}{4}, \frac{3}{4} \right) \alpha \Nc
+ 33600 \Phi_1\left( \frac{3}{4}, \frac{3}{4} \right) \alpha \Nf
- 153000 \Phi_1\left( \frac{3}{4}, \frac{3}{4} \right) \Nc
\right. \nonumber \\
&& \left. ~~~
- 100800 \Phi_1\left( \frac{3}{4}, \frac{3}{4} \right) \Nf
- 10080 \alpha^4 \Nc
+ 135360 \alpha^3 \Nc
- 293760 \alpha^2 \Nc
\right. \nonumber \\
&& \left. ~~~
- 203840 \alpha \Nc
+ 51200 \alpha \Nf
- 43680 \Nc
- 38400 \Nf \right] \frac{\Nc a^2}{460800} \nonumber \\
&& +~ O(a^3) \\  
\gamma_\psi^{\mbox{$\MOMggggs$}}(a,\alpha) &=&
\left[\Nc^2 - 1\right] \frac{\alpha a}{2 \Nc} \nonumber \\ 
&& + \left[ -~ 2592 \ln \left( \frac{4}{3} \right) \alpha^4 \Nc^4
+ 2592 \ln \left( \frac{4}{3} \right) \alpha^4 \Nc^2
+ 145728 \ln \left( \frac{4}{3} \right) \alpha^3 \Nc^4
\right. \nonumber \\
&& \left. ~~~
- 145728 \ln \left( \frac{4}{3} \right) \alpha^3 \Nc^2
+ 38880 \ln \left( \frac{4}{3} \right) \alpha^2 \Nc^4
- 38880 \ln \left( \frac{4}{3} \right) \alpha^2 \Nc^2
\right. \nonumber \\
&& \left. ~~~
- 662016 \ln \left( \frac{4}{3} \right) \alpha \Nc^4
+ 134400 \ln \left( \frac{4}{3} \right) \alpha \Nc^3 \Nf
+ 662016 \ln \left( \frac{4}{3} \right) \alpha \Nc^2
\right. \nonumber \\
&& \left. ~~~
- 134400 \ln \left( \frac{4}{3} \right) \alpha \Nc \Nf
+ 3159 \Phi_1\left( \frac{9}{16}, \frac{9}{16} \right) \alpha^4 \Nc^4
\right. \nonumber \\
&& \left. ~~~
- 3159 \Phi_1\left( \frac{9}{16}, \frac{9}{16} \right) \alpha^4 \Nc^2
- 24381 \Phi_1\left( \frac{9}{16}, \frac{9}{16} \right) \alpha^3 \Nc^4
\right. \nonumber \\
&& \left. ~~~
+ 24381 \Phi_1\left( \frac{9}{16}, \frac{9}{16} \right) \alpha^3 \Nc^2
+ 40365 \Phi_1\left( \frac{9}{16}, \frac{9}{16} \right) \alpha^2 \Nc^4
\right. \nonumber \\
&& \left. ~~~
- 40365 \Phi_1\left( \frac{9}{16}, \frac{9}{16} \right) \alpha^2 \Nc^2
+ 21657 \Phi_1\left( \frac{9}{16}, \frac{9}{16} \right) \alpha \Nc^4
\right. \nonumber \\
&& \left. ~~~
- 4800 \Phi_1\left( \frac{9}{16}, \frac{9}{16} \right) \alpha \Nc^3 \Nf
- 21657 \Phi_1\left( \frac{9}{16}, \frac{9}{16} \right) \alpha \Nc^2
\right. \nonumber \\
&& \left. ~~~
+ 4800 \Phi_1\left( \frac{9}{16}, \frac{9}{16} \right) \alpha \Nc \Nf
- 16200 \Phi_1\left( \frac{3}{4}, \frac{3}{4} \right) \alpha^3 \Nc^4
\right. \nonumber \\
&& \left. ~~~
+ 16200 \Phi_1\left( \frac{3}{4}, \frac{3}{4} \right) \alpha^3 \Nc^2
+ 18000 \Phi_1\left( \frac{3}{4}, \frac{3}{4} \right) \alpha^2 \Nc^4
\right. \nonumber \\
&& \left. ~~~
- 18000 \Phi_1\left( \frac{3}{4}, \frac{3}{4} \right) \alpha^2 \Nc^2
+ 51000 \Phi_1\left( \frac{3}{4}, \frac{3}{4} \right) \alpha \Nc^4
\right. \nonumber \\
&& \left. ~~~
+ 33600 \Phi_1\left( \frac{3}{4}, \frac{3}{4} \right) \alpha \Nc^3 \Nf
- 51000 \Phi_1\left( \frac{3}{4}, \frac{3}{4} \right) \alpha \Nc^2
\right. \nonumber \\
&& \left. ~~~
- 33600 \Phi_1\left( \frac{3}{4}, \frac{3}{4} \right) \alpha \Nc \Nf
- 10080 \alpha^4 \Nc^4
+ 10080 \alpha^4 \Nc^2
\right. \nonumber \\
&& \left. ~~~
+ 105120 \alpha^3 \Nc^4
- 105120 \alpha^3 \Nc^2
+ 21600 \alpha^2 \Nc^4
- 21600 \alpha^2 \Nc^2
\right. \nonumber \\
&& \left. ~~~
- 139040 \alpha \Nc^4
+ 51200 \alpha \Nc^3 \Nf
+ 139040 \alpha \Nc^2
- 51200 \alpha \Nc \Nf
\right. \nonumber \\
&& \left. ~~~
+ 633600 \Nc^4
- 115200 \Nc^3 \Nf
- 547200 \Nc^2
+ 115200 \Nc \Nf
\right. \nonumber \\
&& \left. ~~~
- 86400 \right] \frac{a^2}{230400 \Nc^2} ~+~ O(a^3) 
\end{eqnarray}
and
\begin{eqnarray}
\beta^{\mbox{$\MOMggggs$}}(a,\alpha) &=&
\left[ - 11 \Nc + 2 \Nf \right] \frac{a^2}{3} \nonumber \\
&& + \left[ -~ 2592 \ln \left( \frac{4}{3} \right) \alpha^4 \Nc^3
+ 108384 \ln \left( \frac{4}{3} \right) \alpha^3 \Nc^3
\right. \nonumber \\
&& \left. ~~~ 
- 3456 \ln \left( \frac{4}{3} \right) \alpha^3 \Nc^2 \Nf
- 408032 \ln \left( \frac{4}{3} \right) \alpha^2 \Nc^3
\right. \nonumber \\
&& \left. ~~~ 
+ 129536 \ln \left( \frac{4}{3} \right) \alpha^2 \Nc^2 \Nf
- 56160 \ln \left( \frac{4}{3} \right) \alpha \Nc^3
\right. \nonumber \\
&& \left. ~~~ 
+ 17280 \ln \left( \frac{4}{3} \right) \alpha \Nc^2 \Nf
+ 3159 \Phi_1\left( \frac{9}{16}, \frac{9}{16} \right) \alpha^4 \Nc^3
\right. \nonumber \\
&& \left. ~~~ 
- 29943 \Phi_1\left( \frac{9}{16}, \frac{9}{16} \right) \alpha^3 \Nc^3
+ 4212 \Phi_1\left( \frac{9}{16}, \frac{9}{16} \right) \alpha^3 \Nc^2 \Nf
\right. \nonumber \\
&& \left. ~~~ 
+ 83889 \Phi_1\left( \frac{9}{16}, \frac{9}{16} \right) \alpha^2 \Nc^3
- 21672 \Phi_1\left( \frac{9}{16}, \frac{9}{16} \right) \alpha^2 \Nc^2 \Nf
\right. \nonumber \\
&& \left. ~~~ 
- 58305 \Phi_1\left( \frac{9}{16}, \frac{9}{16} \right) \alpha \Nc^3
+ 17940 \Phi_1\left( \frac{9}{16}, \frac{9}{16} \right) \alpha \Nc^2 \Nf
\right. \nonumber \\
&& \left. ~~~ 
- 10800 \Phi_1\left( \frac{3}{4}, \frac{3}{4} \right) \alpha^3 \Nc^3
+ 52800 \Phi_1\left( \frac{3}{4}, \frac{3}{4} \right) \alpha^2 \Nc^3
\right. \nonumber \\
&& \left. ~~~ 
- 14400 \Phi_1\left( \frac{3}{4}, \frac{3}{4} \right) \alpha^2 \Nc^2 \Nf
- 26000 \Phi_1\left( \frac{3}{4}, \frac{3}{4} \right) \alpha \Nc^3
\right. \nonumber \\
&& \left. ~~~ 
+ 8000 \Phi_1\left( \frac{3}{4}, \frac{3}{4} \right) \alpha \Nc^2 \Nf
- 10080 \alpha^4 \Nc^3
+ 94560 \alpha^3 \Nc^3
\right. \nonumber \\
&& \left. ~~~ 
- 13440 \alpha^3 \Nc^2 \Nf
- 261280 \alpha^2 \Nc^3
+ 67840 \alpha^2 \Nc^2 \Nf
+ 176800 \alpha \Nc^3
\right. \nonumber \\
&& \left. ~~~ 
- 54400 \alpha \Nc^2 \Nf
- 870400 \Nc^3
+ 332800 \Nc^2 \Nf
- 76800 \Nf \right] \frac{a^3}{76800 \Nc} \nonumber \\
&& +~ O(a^4) ~.
\end{eqnarray}
For more practical purposes the numerical values are beneficial and we have
\begin{eqnarray}
\left. \beta^{\mbox{$\MOMggggs$}}(a,\alpha) \right|_{SU(3)} &=&
\left[ 0.666667 \Nf - 11.000000 \right] a^2 \nonumber \\ 
&& + \left[ -~ 0.008632 \alpha^4 - 0.003836 \alpha^3 \Nf - 0.792230 \alpha^3 
- 0.368726 \alpha^2 \Nf 
\right. \nonumber \\
&& \left. ~~~
+ 6.608703 \alpha^2 + 1.339388 \alpha \Nf 
- 13.059036 \alpha 
\right. \nonumber \\
&& \left. ~~~
+ 12.666667 \Nf - 102.000000 \right] a^3 ~+~ O(a^4) \nonumber \\
\left. \gamma_A^{\mbox{$\MOMggggs$}}(a,\alpha) \right|_{SU(3)} &=&
\left[ 1.500000 \alpha + 0.666667 \Nf - 6.500000 \right] a \nonumber \\ 
&& + \left[ -~ 0.002877 \alpha^4 - 0.001279 \alpha^3 \Nf - 1.527349 \alpha^3 
- 0.684363 \alpha^2 \Nf 
\right. \nonumber \\
&& \left. ~~~
+ 8.561163 \alpha^2 + 3.535793 \alpha \Nf 
- 27.533402 \alpha + 0.976180 \Nf^2 
\right. \nonumber \\
&& \left. ~~~
- 3.284136 \Nf - 15.652749 \right] a^2 ~+~ O(a^3) \nonumber \\
\left. \gamma_c^{\mbox{$\MOMggggs$}}(a,\alpha) \right|_{SU(3)} &=&
\left[ 0.750000 \alpha - 2.250000 \right] a \nonumber \\ 
&& + \left[ -~ 0.001439 \alpha^4 + 0.359407 \alpha^3 + 3.254037 \alpha^2 
+ 1.098202 \alpha \Nf 
\right. \nonumber \\
&& \left. ~~~
- 15.132618 \alpha - 2.544606 \Nf - 2.475952 \right] a^2 ~+~ O(a^3)
\nonumber \\
\left. \gamma_\psi^{\mbox{$\MOMggggs$}}(a,\alpha) \right|_{SU(3)} &=&
1.333333 \alpha a \nonumber \\ 
&& + \left[ -~ 0.002557 \alpha^4 + 0.631274 \alpha^3 + 7.678777 \alpha^2 
+ 1.952359 \alpha \Nf 
\right. \nonumber \\
&& \left. ~~~
- 3.866103 \alpha - 1.333333 \Nf + 22.333333 \right] a^2 ~+~ O(a^3)
\end{eqnarray}
for the $SU(3)$ colour group. One can see, for instance, that in the Landau 
gauge the usual two loop QCD $\beta$-function of \cite{51,52} emerges for the
$\MOMgggg$ scheme.

\sect{Discussion.}

We have completed the full symmetric point evaluation of the quartic vertex in
QCD by providing the exact decomposition of the vertex into the full tensor
basis. This extends the earlier work of \cite{18} which was motivated by a
different interest. Broadly we have agreement with \cite{18} where there is 
overlap. One consequence of the determination of the full vertex structure is 
that we are able to define a new momentum subtraction scheme in the same class 
as those proposed in \cite{16}. Unlike the $3$-point vertices which the schemes
of \cite{16} were based on, one has first to be careful in organizing the 
colour group tensors. In other words one has to write the vertex function in 
terms of the structure of the original Feynman and a set of completely 
symmetric colour tensors. The latter are natural for the fully symmetric 
momentum configuration. Once the preferred basis has been determined the 
definition of the MOMgggg scheme emerges naturally. Although we have carried 
out a one loop renormalization properties of the renormalization group equation
have allowed us to construct all the {\em two} loop renormalization group 
functions ahead of an explicit two loop computation. Such an extension would 
require a sizeable calculation. Although there are a significantly large number
of graphs, some, \cite{41}, but not all the basic two loop box master integrals
are known analytically at the fully symmetric point. This is not an
insurmountable obstacle as a numerical evaluation can suffice in the interim
much as (\ref{ptmap}) and (\ref{ptmapval}) did for the one loop case. One
reason why such a two loop computation would be of interest, albeit at one
specific momentum point, is that it would give an estimate of the extent that
the two loop corrections are significant for, say, Schwinger-Dyson comparisons.
Though a more general computation of the fully off-shell quartic vertex would
support future Schwinger-Dyson analyses beyond that carried out, for example,
in \cite{28}.

\vspace{1cm}
\noindent
{\bf Acknowledgements.} The author thanks Dr D.J. Broadhurst for valuable 
discussions.

\appendix

\sect{Tensor basis.}

In this appendix we give the complete set of basis tensors for the quartic
vertex. We have
\begin{eqnarray}
{\cal P}_{(1)}^{\mu \nu \sigma \rho}(p,q,r) &=& \eta^{\mu \nu} \eta^{\sigma \rho} 
~~,~~
{\cal P}_{(2)}^{\mu \nu \sigma \rho}(p,q,r) ~=~ \eta^{\mu \sigma} \eta^{\nu \rho} 
~~,~~
{\cal P}_{(3)}^{\mu \nu \sigma \rho}(p,q,r) ~=~ \eta^{\mu \rho} \eta^{\nu \sigma} 
\nonumber \\
{\cal P}_{(4)}^{\mu \nu \sigma \rho}(p,q,r) &=& \eta^{\mu \nu} \frac{p^\sigma p^\rho}{\mu^2} 
~~,~~
{\cal P}_{(5)}^{\mu \nu \sigma \rho}(p,q,r) ~=~ \eta^{\mu \nu} \frac{p^\sigma q^\rho}{\mu^2} 
~~,~~
{\cal P}_{(6)}^{\mu \nu \sigma \rho}(p,q,r) ~=~ \eta^{\mu \nu} \frac{p^\sigma r^\rho}{\mu^2} 
\nonumber \\
{\cal P}_{(7)}^{\mu \nu \sigma \rho}(p,q,r) &=& \eta^{\mu \nu} \frac{q^\sigma p^\rho}{\mu^2} 
~~,~~
{\cal P}_{(8)}^{\mu \nu \sigma \rho}(p,q,r) ~=~ \eta^{\mu \nu} \frac{q^\sigma q^\rho}{\mu^2} 
~~,~~
{\cal P}_{(9)}^{\mu \nu \sigma \rho}(p,q,r) ~=~ \eta^{\mu \nu} \frac{q^\sigma r^\rho}{\mu^2} 
\nonumber \\
{\cal P}_{(10)}^{\mu \nu \sigma \rho}(p,q,r) &=& \eta^{\mu \nu} \frac{r^\sigma p^\rho}{\mu^2} 
~~,~~
{\cal P}_{(11)}^{\mu \nu \sigma \rho}(p,q,r) ~=~ \eta^{\mu \nu} \frac{r^\sigma q^\rho}{\mu^2} 
~~,~~
{\cal P}_{(12)}^{\mu \nu \sigma \rho}(p,q,r) ~=~ \eta^{\mu \nu} \frac{r^\sigma r^\rho}{\mu^2} 
\nonumber \\
{\cal P}_{(13)}^{\mu \nu \sigma \rho}(p,q,r) &=& \eta^{\mu \sigma} \frac{p^\nu p^\rho}{\mu^2} 
~~,~~
{\cal P}_{(14)}^{\mu \nu \sigma \rho}(p,q,r) ~=~ \eta^{\mu \sigma} \frac{p^\nu q^\rho}{\mu^2} 
~~,~~
{\cal P}_{(15)}^{\mu \nu \sigma \rho}(p,q,r) ~=~ \eta^{\mu \sigma} \frac{p^\nu r^\rho}{\mu^2} 
\nonumber \\
{\cal P}_{(16)}^{\mu \nu \sigma \rho}(p,q,r) &=& \eta^{\mu \sigma} \frac{q^\nu p^\rho}{\mu^2} 
~~,~~
{\cal P}_{(17)}^{\mu \nu \sigma \rho}(p,q,r) ~=~ \eta^{\mu \sigma} \frac{q^\nu q^\rho}{\mu^2} 
~~,~~
{\cal P}_{(18)}^{\mu \nu \sigma \rho}(p,q,r) ~=~ \eta^{\mu \sigma} \frac{q^\nu r^\rho}{\mu^2} 
\nonumber \\
{\cal P}_{(19)}^{\mu \nu \sigma \rho}(p,q,r) &=& \eta^{\mu \sigma} \frac{r^\nu p^\rho}{\mu^2} 
~~,~~
{\cal P}_{(20)}^{\mu \nu \sigma \rho}(p,q,r) ~=~ \eta^{\mu \sigma} \frac{r^\nu q^\rho}{\mu^2} 
~~,~~
{\cal P}_{(21)}^{\mu \nu \sigma \rho}(p,q,r) ~=~ \eta^{\mu \sigma} \frac{r^\nu r^\rho}{\mu^2} 
\nonumber \\
{\cal P}_{(22)}^{\mu \nu \sigma \rho}(p,q,r) &=& \eta^{\mu \rho} \frac{p^\sigma p^\nu}{\mu^2} 
~~,~~
{\cal P}_{(23)}^{\mu \nu \sigma \rho}(p,q,r) ~=~ \eta^{\mu \rho} \frac{p^\sigma q^\nu}{\mu^2} 
~~,~~
{\cal P}_{(24)}^{\mu \nu \sigma \rho}(p,q,r) ~=~ \eta^{\mu \rho} \frac{p^\sigma r^\nu}{\mu^2} 
\nonumber \\
{\cal P}_{(25)}^{\mu \nu \sigma \rho}(p,q,r) &=& \eta^{\mu \rho} \frac{q^\sigma p^\nu}{\mu^2} 
~~,~~
{\cal P}_{(26)}^{\mu \nu \sigma \rho}(p,q,r) ~=~ \eta^{\mu \rho} \frac{q^\sigma q^\nu}{\mu^2} 
~~,~~\
{\cal P}_{(27)}^{\mu \nu \sigma \rho}(p,q,r) ~=~ \eta^{\mu \rho} \frac{q^\sigma r^\nu}{\mu^2} 
\nonumber \\
{\cal P}_{(28)}^{\mu \nu \sigma \rho}(p,q,r) &=& \eta^{\mu \rho} \frac{r^\sigma p^\nu}{\mu^2} 
~~,~~
{\cal P}_{(29)}^{\mu \nu \sigma \rho}(p,q,r) ~=~ \eta^{\mu \rho} \frac{r^\sigma q^\nu}{\mu^2} 
~~,~~\
{\cal P}_{(30)}^{\mu \nu \sigma \rho}(p,q,r) ~=~ \eta^{\mu \rho} \frac{r^\sigma r^\nu}{\mu^2} 
\nonumber \\
{\cal P}_{(31)}^{\mu \nu \sigma \rho}(p,q,r) &=& \eta^{\nu \sigma} \frac{p^\mu p^\rho}{\mu^2} 
~~,~~
{\cal P}_{(32)}^{\mu \nu \sigma \rho}(p,q,r) ~=~ \eta^{\nu \sigma} \frac{p^\mu q^\rho}{\mu^2} 
~~,~~\
{\cal P}_{(33)}^{\mu \nu \sigma \rho}(p,q,r) ~=~ \eta^{\nu \sigma} \frac{p^\mu r^\rho}{\mu^2} 
\nonumber \\
{\cal P}_{(34)}^{\mu \nu \sigma \rho}(p,q,r) &=& \eta^{\nu \sigma} \frac{q^\mu p^\rho}{\mu^2} 
~~,~~
{\cal P}_{(35)}^{\mu \nu \sigma \rho}(p,q,r) ~=~ \eta^{\nu \sigma} \frac{q^\mu q^\rho}{\mu^2} 
~~,~~\
{\cal P}_{(36)}^{\mu \nu \sigma \rho}(p,q,r) ~=~ \eta^{\nu \sigma} \frac{q^\mu r^\rho}{\mu^2} 
\nonumber \\
{\cal P}_{(37)}^{\mu \nu \sigma \rho}(p,q,r) &=& \eta^{\nu \sigma} \frac{r^\mu p^\rho}{\mu^2} 
~~,~~
{\cal P}_{(38)}^{\mu \nu \sigma \rho}(p,q,r) ~=~ \eta^{\nu \sigma} \frac{r^\mu q^\rho}{\mu^2} 
~~,~~\
{\cal P}_{(39)}^{\mu \nu \sigma \rho}(p,q,r) ~=~ \eta^{\nu \sigma} \frac{r^\mu r^\rho}{\mu^2} 
\nonumber \\
{\cal P}_{(40)}^{\mu \nu \sigma \rho}(p,q,r) &=& \eta^{\nu \rho} \frac{p^\mu p^\sigma}{\mu^2} 
~~,~~
{\cal P}_{(41)}^{\mu \nu \sigma \rho}(p,q,r) ~=~ \eta^{\nu \rho} \frac{p^\mu q^\sigma}{\mu^2} 
~~,~~\
{\cal P}_{(42)}^{\mu \nu \sigma \rho}(p,q,r) ~=~ \eta^{\nu \rho} \frac{p^\mu r^\sigma}{\mu^2} 
\nonumber \\
{\cal P}_{(43)}^{\mu \nu \sigma \rho}(p,q,r) &=& \eta^{\nu \rho} \frac{q^\mu p^\sigma}{\mu^2} 
~~,~~
{\cal P}_{(44)}^{\mu \nu \sigma \rho}(p,q,r) ~=~ \eta^{\nu \rho} \frac{q^\mu q^\sigma}{\mu^2} 
~~,~~\
{\cal P}_{(45)}^{\mu \nu \sigma \rho}(p,q,r) ~=~ \eta^{\nu \rho} \frac{q^\mu r^\sigma}{\mu^2} 
\nonumber \\
{\cal P}_{(46)}^{\mu \nu \sigma \rho}(p,q,r) &=& \eta^{\nu \rho} \frac{r^\mu p^\sigma}{\mu^2} 
~~,~~
{\cal P}_{(47)}^{\mu \nu \sigma \rho}(p,q,r) ~=~ \eta^{\nu \rho} \frac{r^\mu q^\sigma}{\mu^2} 
~~,~~\
{\cal P}_{(48)}^{\mu \nu \sigma \rho}(p,q,r) ~=~ \eta^{\nu \rho} \frac{r^\mu r^\sigma}{\mu^2} 
\nonumber \\
{\cal P}_{(49)}^{\mu \nu \sigma \rho}(p,q,r) &=& \eta^{\sigma \rho} \frac{p^\mu p^\nu}{\mu^2} 
~~,~~
{\cal P}_{(50)}^{\mu \nu \sigma \rho}(p,q,r) ~=~ \eta^{\sigma \rho} \frac{p^\mu q^\nu}{\mu^2} 
~~,~~\
{\cal P}_{(51)}^{\mu \nu \sigma \rho}(p,q,r) ~=~ \eta^{\sigma \rho} \frac{p^\mu r^\nu}{\mu^2} 
\nonumber \\
{\cal P}_{(52)}^{\mu \nu \sigma \rho}(p,q,r) &=& \eta^{\sigma \rho} \frac{q^\mu p^\nu}{\mu^2} 
~~,~~
{\cal P}_{(53)}^{\mu \nu \sigma \rho}(p,q,r) ~=~ \eta^{\sigma \rho} \frac{q^\mu q^\nu}{\mu^2} 
~~,~~\
{\cal P}_{(54)}^{\mu \nu \sigma \rho}(p,q,r) ~=~ \eta^{\sigma \rho} \frac{q^\mu r^\nu}{\mu^2} 
\nonumber \\
{\cal P}_{(55)}^{\mu \nu \sigma \rho}(p,q,r) &=& \eta^{\sigma \rho} \frac{r^\mu p^\nu}{\mu^2} 
~~,~~
{\cal P}_{(56)}^{\mu \nu \sigma \rho}(p,q,r) ~=~ \eta^{\sigma \rho} \frac{r^\mu q^\nu}{\mu^2} 
~~,~~\
{\cal P}_{(57)}^{\mu \nu \sigma \rho}(p,q,r) ~=~ \eta^{\sigma \rho} \frac{r^\mu r^\nu}{\mu^2} 
\nonumber \\
{\cal P}_{(58)}^{\mu \nu \sigma \rho}(p,q,r) &=& \frac{p^\mu p^\nu p^\sigma p^\rho}{\mu^4} 
~~,~~
{\cal P}_{(59)}^{\mu \nu \sigma \rho}(p,q,r) ~=~ \frac{q^\mu q^\nu q^\sigma q^\rho}{\mu^4} 
~~,~~\
{\cal P}_{(60)}^{\mu \nu \sigma \rho}(p,q,r) ~=~ \frac{r^\mu r^\nu r^\sigma r^\rho}{\mu^4} 
\nonumber \\
{\cal P}_{(61)}^{\mu \nu \sigma \rho}(p,q,r) &=& \frac{p^\mu p^\nu p^\sigma q^\rho}{\mu^4} 
~~,~~
{\cal P}_{(62)}^{\mu \nu \sigma \rho}(p,q,r) ~=~ \frac{p^\mu p^\nu q^\sigma p^\rho}{\mu^4} 
~~,~~\
{\cal P}_{(63)}^{\mu \nu \sigma \rho}(p,q,r) ~=~ \frac{p^\mu q^\nu p^\sigma p^\rho}{\mu^4} 
\nonumber \\
{\cal P}_{(64)}^{\mu \nu \sigma \rho}(p,q,r) &=& \frac{q^\mu p^\nu p^\sigma p^\rho}{\mu^4} 
~~,~~
{\cal P}_{(65)}^{\mu \nu \sigma \rho}(p,q,r) ~=~ \frac{p^\mu p^\nu p^\sigma r^\rho}{\mu^4} 
~~,~~\
{\cal P}_{(66)}^{\mu \nu \sigma \rho}(p,q,r) ~=~ \frac{p^\mu p^\nu r^\sigma p^\rho}{\mu^4} 
\nonumber \\
{\cal P}_{(67)}^{\mu \nu \sigma \rho}(p,q,r) &=& \frac{p^\mu r^\nu p^\sigma p^\rho}{\mu^4} 
~~,~~
{\cal P}_{(68)}^{\mu \nu \sigma \rho}(p,q,r) ~=~ \frac{r^\mu p^\nu p^\sigma p^\rho}{\mu^4} 
~~,~~\
{\cal P}_{(69)}^{\mu \nu \sigma \rho}(p,q,r) ~=~ \frac{q^\mu q^\nu q^\sigma r^\rho}{\mu^4} 
\nonumber \\
{\cal P}_{(70)}^{\mu \nu \sigma \rho}(p,q,r) &=& \frac{q^\mu q^\nu r^\sigma q^\rho}{\mu^4} 
~~,~~
{\cal P}_{(71)}^{\mu \nu \sigma \rho}(p,q,r) ~=~ \frac{q^\mu r^\nu q^\sigma q^\rho}{\mu^4} 
~~,~~\
{\cal P}_{(72)}^{\mu \nu \sigma \rho}(p,q,r) ~=~ \frac{r^\mu q^\nu q^\sigma q^\rho}{\mu^4} 
\nonumber \\
{\cal P}_{(73)}^{\mu \nu \sigma \rho}(p,q,r) &=& \frac{r^\mu r^\nu r^\sigma q^\rho}{\mu^4} 
~~,~~
{\cal P}_{(74)}^{\mu \nu \sigma \rho}(p,q,r) ~=~ \frac{r^\mu r^\nu q^\sigma r^\rho}{\mu^4} 
~~,~~\
{\cal P}_{(75)}^{\mu \nu \sigma \rho}(p,q,r) ~=~ \frac{r^\mu q^\nu r^\sigma r^\rho}{\mu^4} 
\nonumber \\
{\cal P}_{(76)}^{\mu \nu \sigma \rho}(p,q,r) &=& \frac{q^\mu r^\nu r^\sigma r^\rho}{\mu^4} 
~~,~~
{\cal P}_{(77)}^{\mu \nu \sigma \rho}(p,q,r) ~=~ \frac{r^\mu r^\nu r^\sigma p^\rho}{\mu^4} 
~~,~~\
{\cal P}_{(78)}^{\mu \nu \sigma \rho}(p,q,r) ~=~ \frac{r^\mu r^\nu p^\sigma r^\rho}{\mu^4} 
\nonumber \\
{\cal P}_{(79)}^{\mu \nu \sigma \rho}(p,q,r) &=& \frac{r^\mu p^\nu r^\sigma r^\rho}{\mu^4} 
~~,~~
{\cal P}_{(80)}^{\mu \nu \sigma \rho}(p,q,r) ~=~ \frac{p^\mu r^\nu r^\sigma r^\rho}{\mu^4} 
~~,~~\
{\cal P}_{(81)}^{\mu \nu \sigma \rho}(p,q,r) ~=~ \frac{q^\mu q^\nu q^\sigma p^\rho}{\mu^4} 
\nonumber \\
{\cal P}_{(82)}^{\mu \nu \sigma \rho}(p,q,r) &=& \frac{q^\mu q^\nu p^\sigma q^\rho}{\mu^4} 
~~,~~
{\cal P}_{(83)}^{\mu \nu \sigma \rho}(p,q,r) ~=~ \frac{q^\mu p^\nu q^\sigma q^\rho}{\mu^4} 
~~,~~\
{\cal P}_{(84)}^{\mu \nu \sigma \rho}(p,q,r) ~=~ \frac{p^\mu q^\nu q^\sigma q^\rho}{\mu^4} 
\nonumber \\
{\cal P}_{(85)}^{\mu \nu \sigma \rho}(p,q,r) &=& \frac{p^\mu p^\nu q^\sigma q^\rho}{\mu^4} 
~~,~~
{\cal P}_{(86)}^{\mu \nu \sigma \rho}(p,q,r) ~=~ \frac{p^\mu q^\nu p^\sigma q^\rho}{\mu^4} 
~~,~~\
{\cal P}_{(87)}^{\mu \nu \sigma \rho}(p,q,r) ~=~ \frac{q^\mu p^\nu p^\sigma q^\rho}{\mu^4} 
\nonumber \\
{\cal P}_{(88)}^{\mu \nu \sigma \rho}(p,q,r) &=& \frac{p^\mu q^\nu q^\sigma p^\rho}{\mu^4} 
~~,~~
{\cal P}_{(89)}^{\mu \nu \sigma \rho}(p,q,r) ~=~ \frac{q^\mu p^\nu q^\sigma p^\rho}{\mu^4} 
~~,~~\
{\cal P}_{(90)}^{\mu \nu \sigma \rho}(p,q,r) ~=~ \frac{q^\mu q^\nu p^\sigma p^\rho}{\mu^4} 
\nonumber \\
{\cal P}_{(91)}^{\mu \nu \sigma \rho}(p,q,r) &=& \frac{p^\mu p^\nu r^\sigma r^\rho}{\mu^4} 
~~,~~
{\cal P}_{(92)}^{\mu \nu \sigma \rho}(p,q,r) ~=~ \frac{p^\mu r^\nu p^\sigma r^\rho}{\mu^4} 
~~,~~\
{\cal P}_{(93)}^{\mu \nu \sigma \rho}(p,q,r) ~=~ \frac{r^\mu p^\nu p^\sigma r^\rho}{\mu^4} 
\nonumber \\
{\cal P}_{(94)}^{\mu \nu \sigma \rho}(p,q,r) &=& \frac{p^\mu r^\nu r^\sigma p^\rho}{\mu^4} 
~~,~~
{\cal P}_{(95)}^{\mu \nu \sigma \rho}(p,q,r) ~=~ \frac{r^\mu p^\nu r^\sigma p^\rho}{\mu^4} 
~~,~~\
{\cal P}_{(96)}^{\mu \nu \sigma \rho}(p,q,r) ~=~ \frac{r^\mu r^\nu p^\sigma p^\rho}{\mu^4} 
\nonumber \\
{\cal P}_{(97)}^{\mu \nu \sigma \rho}(p,q,r) &=& \frac{q^\mu q^\nu r^\sigma r^\rho}{\mu^4} 
~~,~~
{\cal P}_{(98)}^{\mu \nu \sigma \rho}(p,q,r) ~=~ \frac{q^\mu r^\nu q^\sigma r^\rho}{\mu^4} 
~~,~~\
{\cal P}_{(99)}^{\mu \nu \sigma \rho}(p,q,r) ~=~ \frac{r^\mu q^\nu q^\sigma r^\rho}{\mu^4} 
\nonumber \\
{\cal P}_{(100)}^{\mu \nu \sigma \rho}(p,q,r) &=& \frac{q^\mu r^\nu r^\sigma q^\rho}{\mu^4} 
~~,~~
{\cal P}_{(101)}^{\mu \nu \sigma \rho}(p,q,r) ~=~ \frac{r^\mu q^\nu r^\sigma q^\rho}{\mu^4} 
~~,~~\
{\cal P}_{(102)}^{\mu \nu \sigma \rho}(p,q,r) ~=~ \frac{r^\mu r^\nu q^\sigma q^\rho}{\mu^4} 
\nonumber \\
{\cal P}_{(103)}^{\mu \nu \sigma \rho}(p,q,r) &=& \frac{p^\mu p^\nu q^\sigma r^\rho}{\mu^4} 
~~,~~
{\cal P}_{(104)}^{\mu \nu \sigma \rho}(p,q,r) ~=~ \frac{p^\mu q^\nu p^\sigma r^\rho}{\mu^4} 
~~,~~\
{\cal P}_{(105)}^{\mu \nu \sigma \rho}(p,q,r) ~=~ \frac{q^\mu p^\nu p^\sigma r^\rho}{\mu^4} 
\nonumber \\
{\cal P}_{(106)}^{\mu \nu \sigma \rho}(p,q,r) &=& \frac{p^\mu q^\nu r^\sigma p^\rho}{\mu^4} 
~~,~~
{\cal P}_{(107)}^{\mu \nu \sigma \rho}(p,q,r) ~=~ \frac{q^\mu p^\nu r^\sigma p^\rho}{\mu^4} 
~~,~~\
{\cal P}_{(108)}^{\mu \nu \sigma \rho}(p,q,r) ~=~ \frac{q^\mu r^\nu p^\sigma p^\rho}{\mu^4} 
\nonumber \\
{\cal P}_{(109)}^{\mu \nu \sigma \rho}(p,q,r) &=& \frac{p^\mu p^\nu r^\sigma q^\rho}{\mu^4} 
~~,~~
{\cal P}_{(110)}^{\mu \nu \sigma \rho}(p,q,r) ~=~ \frac{p^\mu r^\nu p^\sigma q^\rho}{\mu^4} 
~~,~~\
{\cal P}_{(111)}^{\mu \nu \sigma \rho}(p,q,r) ~=~ \frac{r^\mu p^\nu p^\sigma q^\rho}{\mu^4} 
\nonumber \\
{\cal P}_{(112)}^{\mu \nu \sigma \rho}(p,q,r) &=& \frac{p^\mu r^\nu q^\sigma p^\rho}{\mu^4} 
~~,~~
{\cal P}_{(113)}^{\mu \nu \sigma \rho}(p,q,r) ~=~ \frac{r^\mu p^\nu q^\sigma p^\rho}{\mu^4} 
~~,~~\
{\cal P}_{(114)}^{\mu \nu \sigma \rho}(p,q,r) ~=~ \frac{r^\mu q^\nu p^\sigma p^\rho}{\mu^4} 
\nonumber \\
{\cal P}_{(115)}^{\mu \nu \sigma \rho}(p,q,r) &=& \frac{q^\mu q^\nu r^\sigma p^\rho}{\mu^4} 
~~,~~
{\cal P}_{(116)}^{\mu \nu \sigma \rho}(p,q,r) ~=~ \frac{q^\mu r^\nu q^\sigma p^\rho}{\mu^4} 
~~,~~\
{\cal P}_{(117)}^{\mu \nu \sigma \rho}(p,q,r) ~=~ \frac{r^\mu q^\nu q^\sigma p^\rho}{\mu^4} 
\nonumber \\
{\cal P}_{(118)}^{\mu \nu \sigma \rho}(p,q,r) &=& \frac{q^\mu r^\nu p^\sigma q^\rho}{\mu^4} 
~~,~~
{\cal P}_{(119)}^{\mu \nu \sigma \rho}(p,q,r) ~=~ \frac{r^\mu q^\nu p^\sigma q^\rho}{\mu^4} 
~~,~~\
{\cal P}_{(120)}^{\mu \nu \sigma \rho}(p,q,r) ~=~ \frac{r^\mu p^\nu q^\sigma q^\rho}{\mu^4} 
\nonumber \\
{\cal P}_{(121)}^{\mu \nu \sigma \rho}(p,q,r) &=& \frac{q^\mu q^\nu p^\sigma r^\rho}{\mu^4} 
~~,~~
{\cal P}_{(122)}^{\mu \nu \sigma \rho}(p,q,r) ~=~ \frac{q^\mu p^\nu q^\sigma r^\rho}{\mu^4} 
~~,~~\
{\cal P}_{(123)}^{\mu \nu \sigma \rho}(p,q,r) ~=~ \frac{p^\mu q^\nu q^\sigma r^\rho}{\mu^4} 
\nonumber \\
{\cal P}_{(124)}^{\mu \nu \sigma \rho}(p,q,r) &=& \frac{q^\mu p^\nu r^\sigma q^\rho}{\mu^4} 
~~,~~
{\cal P}_{(125)}^{\mu \nu \sigma \rho}(p,q,r) ~=~ \frac{p^\mu q^\nu r^\sigma q^\rho}{\mu^4} 
~~,~~\
{\cal P}_{(126)}^{\mu \nu \sigma \rho}(p,q,r) ~=~ \frac{p^\mu r^\nu q^\sigma q^\rho}{\mu^4} 
\nonumber \\
{\cal P}_{(127)}^{\mu \nu \sigma \rho}(p,q,r) &=& \frac{r^\mu r^\nu p^\sigma q^\rho}{\mu^4} 
~~,~~
{\cal P}_{(128)}^{\mu \nu \sigma \rho}(p,q,r) ~=~ \frac{r^\mu p^\nu r^\sigma q^\rho}{\mu^4} 
~~,~~\
{\cal P}_{(129)}^{\mu \nu \sigma \rho}(p,q,r) ~=~ \frac{p^\mu r^\nu r^\sigma q^\rho}{\mu^4} 
\nonumber \\
{\cal P}_{(130)}^{\mu \nu \sigma \rho}(p,q,r) &=& \frac{r^\mu p^\nu q^\sigma r^\rho}{\mu^4} 
~~,~~
{\cal P}_{(131)}^{\mu \nu \sigma \rho}(p,q,r) ~=~ \frac{p^\mu r^\nu q^\sigma r^\rho}{\mu^4} 
~~,~~\
{\cal P}_{(132)}^{\mu \nu \sigma \rho}(p,q,r) ~=~ \frac{p^\mu q^\nu r^\sigma r^\rho}{\mu^4} 
\nonumber \\
{\cal P}_{(133)}^{\mu \nu \sigma \rho}(p,q,r) &=& \frac{r^\mu r^\nu q^\sigma p^\rho}{\mu^4} 
~~,~~
{\cal P}_{(134)}^{\mu \nu \sigma \rho}(p,q,r) ~=~ \frac{r^\mu q^\nu r^\sigma p^\rho}{\mu^4} 
~~,~~\
{\cal P}_{(135)}^{\mu \nu \sigma \rho}(p,q,r) ~=~ \frac{q^\mu r^\nu r^\sigma p^\rho}{\mu^4} 
\nonumber \\
{\cal P}_{(136)}^{\mu \nu \sigma \rho}(p,q,r) &=& \frac{r^\mu q^\nu p^\sigma r^\rho}{\mu^4} 
~~,~~
{\cal P}_{(137)}^{\mu \nu \sigma \rho}(p,q,r) ~=~ \frac{q^\mu r^\nu p^\sigma r^\rho}{\mu^4} 
~~,~~\
{\cal P}_{(138)}^{\mu \nu \sigma \rho}(p,q,r) ~=~ \frac{q^\mu p^\nu r^\sigma r^\rho}{\mu^4} ~.
\nonumber \\
\end{eqnarray}
In previous similar and related work we have at this point given the 
corresponding projection matrix, ${\cal M}^i_{kl}$, which allows one to 
determine each channel of the Green's function. However, due to the size of 
this matrix we have relegated the explicit matrix elements to the data file.

\end{document}